\documentclass[twocolumn,showpacs,preprintnumbers,amsmath,amssymb,aps,prb]{revtex4}
\usepackage{graphicx}
\begin{document}

\title{
Transverse ac Driven and Geometric Ratchet Effects for Vortices in Conformal Crystal Pinning Arrays 
} 
\author{
C. Reichhardt 
and C. J. Olson Reichhardt
} 
\address{
Theoretical Division,
Los Alamos National Laboratory, Los Alamos, New Mexico 87545 USA 
} 

\begin{abstract}
A conformal pinning array is created by taking a conformal 
transformation of a uniform hexagonal lattice 
to create
a new structure in which the six-fold ordering of the original lattice is preserved 
but which has a spatial gradient in the pinning site density.
With a series of conformal arrays it is possible to create asymmetric
substrates, and
it was previously shown that when an ac drive is applied parallel to
the asymmetry direction, 
a pronounced ratchet effect occurs with a net dc flow of vortices in the same direction
as the ac drive.
Here we show that when the ac drive is applied perpendicular to the
substrate asymmetry
direction, it is possible to realize a transverse ratchet effect where a net dc flow of 
vortices is generated perpendicular to the
ac drive.
The conformal transverse ratchet effect is distinct from previous
versions of transverse ratchets in that 
it occurs due to the generation of non-Gaussian transverse vortex velocity
fluctuations by the plastic motion of vortices, so that the 
system behaves as a noise correlation ratchet.
The transverse ratchet
effect is much more pronounced in the conformal arrays
than in random gradient arrays, and is absent in square 
gradient arrays due the different nature of the
vortex flow in  each geometry.
We show that a series of reversals can occur in the transverse ratchet effect
due to changes in the vortex 
flow across the pinning gradient as a function of
vortex filling, pinning strength, and ac amplitude.    
We also consider the case where a dc drive applied perpendicular to the substrate
asymmetry direction generates a net flow of 
vortices perpendicular to the dc drive, producing what is known as a
geometric or drift ratchet that again arises due 
to non-Gaussian dynamically generated fluctuations.
The drift ratchet is more efficient than the ac driven ratchet
and also exhibits a series of reversals for
varied parameters.  Our results should be general to a wide class of systems
undergoing nonequilibrium dynamics on conformal substrates, such as colloidal particles
on optical traps.      
\end{abstract}
\maketitle

\section{Introduction}
When a particle   
interacts with an asymmetric substrate under an ac drive,
it is possible to realize a ratchet effect 
in which a net dc motion of the particle can occur.  
If the ac drive is in the form of a periodic external force,
the system is characterized as a  
rocking ratchet, while
if the particles are undergoing thermal agitation and
the substrate is periodically flashed on and off,
the system is known as a flashing ratchet \cite{1,2}. 
It is also possible to generate dc motion of a particle in an asymmetric substrate
in the absence
of periodic driving when the 
noise fluctuations experienced by the particle are not white but have an
additional time correlation.
Such systems are known as correlation ratchets \cite{3,4,5,6}.
Another
type of ratchet system, the geometric
or drift ratchet,
arises when dc driven particles move through an assembly
of asymmetric obstacles.
The interactions of the particles with the obstacles produces
an additional drift of particles in the
direction perpendicular to the dc drive \cite{7,8,9,10,11}.
One system in which a remarkably rich variety of rocking ratchet effects has been realized is
vortices in type-II superconductors 
interacting with some form of asymmetric
substrate where an ac driving force is applied along the  direction
of the substrate asymmetry
\cite{12,12N,14,15,16,17,18,19,20,21,22,23,24}.
Such asymmetric pinning arrays include periodic
one-dimensional (1D) asymmetric saw-tooth thickness modulations 
\cite{12,19,23,25,26},
channels with funnel shapes \cite{12N,20,21,27,28},
and arrays of pinning sites in which the
individual pinning sites have an intrinsic
asymmetry \cite{15,17,18,22,24,29,30}. 
In the case where the vortex-vortex interactions are weak,
such as at low magnetic fields, the net dc flow 
is in the easy direction of the substrate asymmetry,
but when collective vortex interactions
come into play it is possible to observe reversals of the ratchet effect 
in which the net dc vortex motion switches to follow the hard direction
of the substrate asymmetry, and there may be
multiple ratchet reversals as a function of vortex density and ac driving amplitudes 
\cite{15,18,19,25,26,33,34,35,36,37}.
It is also possible to create transverse vortex ratchets where a net dc flow of vortices 
occurs that is perpendicular to the applied ac driving force \cite{31,38,36a,39,40}.
Such transverse ratchets have been realized
for vortices interacting with arrays of triangular pins
or other pinning site shapes that have an intrinsic asymmetry.
When vortices interact with such pinning sites, they are  partially
deflected in the direction perpendicular to the ac force  
\cite{39,40}.

Vortex ratchet pinning geometries can be created using individual pinning
sites that are symmetric by introducing 
a periodic spatial gradient in the pinning site density
\cite{14,41,42,43}
or a gradient in the size of the pinning
sites \cite{44}, and then applying an ac driving force along the direction of the gradient.
Gradient ratchets have been studied for
randomly arranged pinning with a spatial gradient \cite{41}
and periodically arranged pinning arrays with gradients \cite{41,42,43,44}.
A new type of pinning geometry called a conformal crystal was recently
proposed \cite{47}.  It is constructed by 
applying a conformal transformation to a uniform hexagonal array, resulting in
a structure 
where each pinning site has six neighbors but there is a spatial
gradient in the pinning site density  \cite{45,46}. 
Simulations indicate that the pinning effectiveness for a fixed number of pinning
sites is maximized for the conformal arrays
compared to uniform random arrays or a random arrangement of pinning
sites with a gradient \cite{47}.
The conformal array also produces more effective pinning
than uniform periodic pinning arrays except for fields at or very near
integer matching fields. 
This optimal pinning by conformal arrays
was subsequently confirmed in experiments \cite{48,49}.
Other studies have also shown that
non-conformal gradient pinning arrays
produce enhanced pinning compared to uniform arrays \cite{50,M}.
Simulations
demonstrate that the hexagonal conformal arrays
exhibit stronger pinning than arrays constructed by  conformally
transforming square or quasiperiodic arrays, while  
square pinning arrays containing 1D spatial density gradients
have enhanced pinning compared to the hexagonal conformal arrays
at certain magnetic field fillings due to commensuration effects \cite{51}.

Vortices interacting with a series of conformal pinning arrays 
can exhibit a ratchet effect 
when an external ac drive is applied along the substrate asymmetry direction,
and multiple reversals
in the ratchet effect can occur as a function of vortex density,
pinning force, and drive amplitude due to changes in the
vortex flow patterns, as shown in Ref.~\cite{52}.
In general, the ratchet effect is most pronounced for the conformal
pinning arrays compared
to random gradient or square gradient pinning arrays;
however, at low magnetic fields the ratchet effect is strongest for the square
gradient array
due to a 1D channeling of vortices along
the symmetry direction of the square array.

For conformal or other gradient pinning arrays, 
it has not been considered
whether a transverse ratchet effect could occur when the
ac drive is applied {\it perpendicular} to the substrate asymmetry direction. 
Since the individual pinning sites are symmetric in these arrays,
it might be expected that transverse ratchet effects would not occur.
Here we show that 
a pronounced transverse vortex ratchet effect appears in conformal pinning arrays,
and that the mechanism responsible
for this effect 
differs from that which produces
the transverse ratchet effects found for asymmetrically shaped pins or obstacles.
For the conformal
array, the transverse ratchet effect results when plastic vortex motion 
generates strong non-Gaussian velocity fluctuations  
both parallel and perpendicular to the ac driving direction.
Due to the pinning gradient, the 
fluctuations can be stronger in the low pinning density portions
of the sample, creating an effective gradient in the
size of the fluctuations
and producing a dynamical thermophoresis phenomenon.
The effect can also be viewed as a realization of a noise correlation ratchet.
Noise correlation ratchets were first proposed by  
Magnasco \cite{3} and Doering {\it et al.} \cite{4}, who introduced
different types of noise correlations directly into the fluctuating noise term governing
the equation of motion of a particle placed on an asymmetric substrate.
In the conformal pinning array system that we consider, there is no 
added stochastic noise term in the vortex equation of motion,
so the correlated velocity fluctuations are dynamically generated 
by the plastic motion of the vortices.
The onset of plastic flow in vortex systems has
been demonstrated both in experiments \cite{53} and simulations
to generate strong non-Gaussian vortex velocity fluctuations
both parallel and perpendicular 
to the external driving  direction \cite{54,55}.   
We find that transverse ratchet effects do not occur in square gradient arrays since the
vortex trajectories are nearly one-dimensional in the direction parallel to the
ac drive, so the transverse fluctuations 
are too weak to induce transverse ratchet motion.
We show that it is possible to realize 
geometric or drift ratchet effects in the conformal and random gradient arrays
when  a dc drive is applied perpendicular to the asymmetry direction and a net vortex drift 
arises that is perpendicular to the dc drive.
Geometrical
ratchets have been studied for the dc flow of particles through
periodic arrays of asymmetric objects, and can arise even in the limit 
of a single particle \cite{7,8,9,10,11}. In our
system the pinning sites are symmetric but a geometric ratchet effect
occurs due to the dynamically generated fluctuations
from the plastic flow of vortices.  In the single particle limit where
collective plastic flow is lost, the transverse or geometric ratchet effect is also absent.  
The geometric ratchet we study has similarities 
to the geometric ratchet effect
proposed by Kolton \cite{56}
for particles in a two-dimensional (2D) system moving    
over a periodic 1D asymmetric substrate containing additional random pinning sites
that create nonequilibrium fluctuations, which 
result in a drift of particles perpendicular to the dc drive direction.
We also find that there can be a series of
reversals in both the ac and dc driven transverse ratchet effects which is
unique among transverse and drift ratchet systems that
have been previously studied.
Finally, our results should be general to a variety of  of other systems which can be described
as an assembly of interacting particles moving over a conformal pinning array,
such as colloids on optical trap arrays \cite{57}.  

\section{Simulation and System}   
We consider a 2D simulation geometry with periodic boundary conditions
in the $x$ and $y$-directions. 
Within the system we place $N_{v}$ vortices, where the number of
vortices is proportional to the applied magnetic field 
$B$ which is aligned in the ${\bf \hat z}$ direction.
We introduce $N_{p}$ pinning sites to the sample, and denote the field at which
the number of vortices equals the number of pinning sites
as $B_{\phi}$.
The dynamics of an individual vortex $i$ is governed  
by the following overdamped equation of motion:
\begin{equation}  
\eta \frac{d {\bf R}_{i}}{dt} = 
{\bf F}^{vv}_{i} + {\bf F}^{vp}_{i} + {\bf F}^{ac}_{i} + {\bf F}^{dc}_{i}  + {\bf F}^T_i.
\end{equation} 
Here $\eta$ is the damping constant
which is set equal to $1$.
The first term on the right is the repulsive vortex-vortex 
interaction force  
${\bf F}^{vv}_{i} = \sum_{j\neq i}^{N_v}F_{0}K_{1}(R_{ij}/\lambda){\hat {\bf R}}_{ij}$,
where ${\bf R}_{i}$ is the location of vortex $i$,
$K_{1}$ is the modified Bessel function, $\lambda$ is the penetration depth, 
$R_{ij} = |{\bf R}_{i} - {\bf R}_{j}|$,
$ {\hat {\bf R}_{ij}} = ({\bf R}_{i} - {\bf R}_{j})/R_{ij}$, 
$F_{0} = \phi^{2}_{0}/(2\pi\mu_{0}\lambda^3)$, 
$\phi_{0}$ is the flux quantum, and $\mu_{0}$ is the permittivity.    
The initial vortex positions before application of an external driving force 
are obtained by
performing simulated annealing from a high temperature state down to $T = 0.$
The vortex-pinning interaction term 
${\bf F}^{vp}_{i} = \sum^{N_{p}}_{k=1}(F_{p}R^{(p)}_{ik}/r_{p})
\Theta((r_{p} -R^{(p)}_{ik})/\lambda){\hat {\bf R}^{(p)}}_{ik}$,  
where $\Theta$ is the Heaviside step function, 
$r_{p} = 0.25\lambda$ is the pinning radius, $F_{p}$ is the pinning strength, 
${\bf R}_k^{(p)}$ is the location of pinning site $k$,
$R_{ik}^{(p)} = |{\bf R}_{i} - {\bf R}_{k}^{(p)}|$, and
$ {\hat {\bf R}_{ik}^{(p)}} = ({\bf R}_{i} - {\bf R}_{k}^{(p)})/R_{ik}^{(p)}$. 
All forces are measured in units of $F_{0}$ and lengths in units of $\lambda$.

\begin{figure}
\includegraphics[width=3.5in]{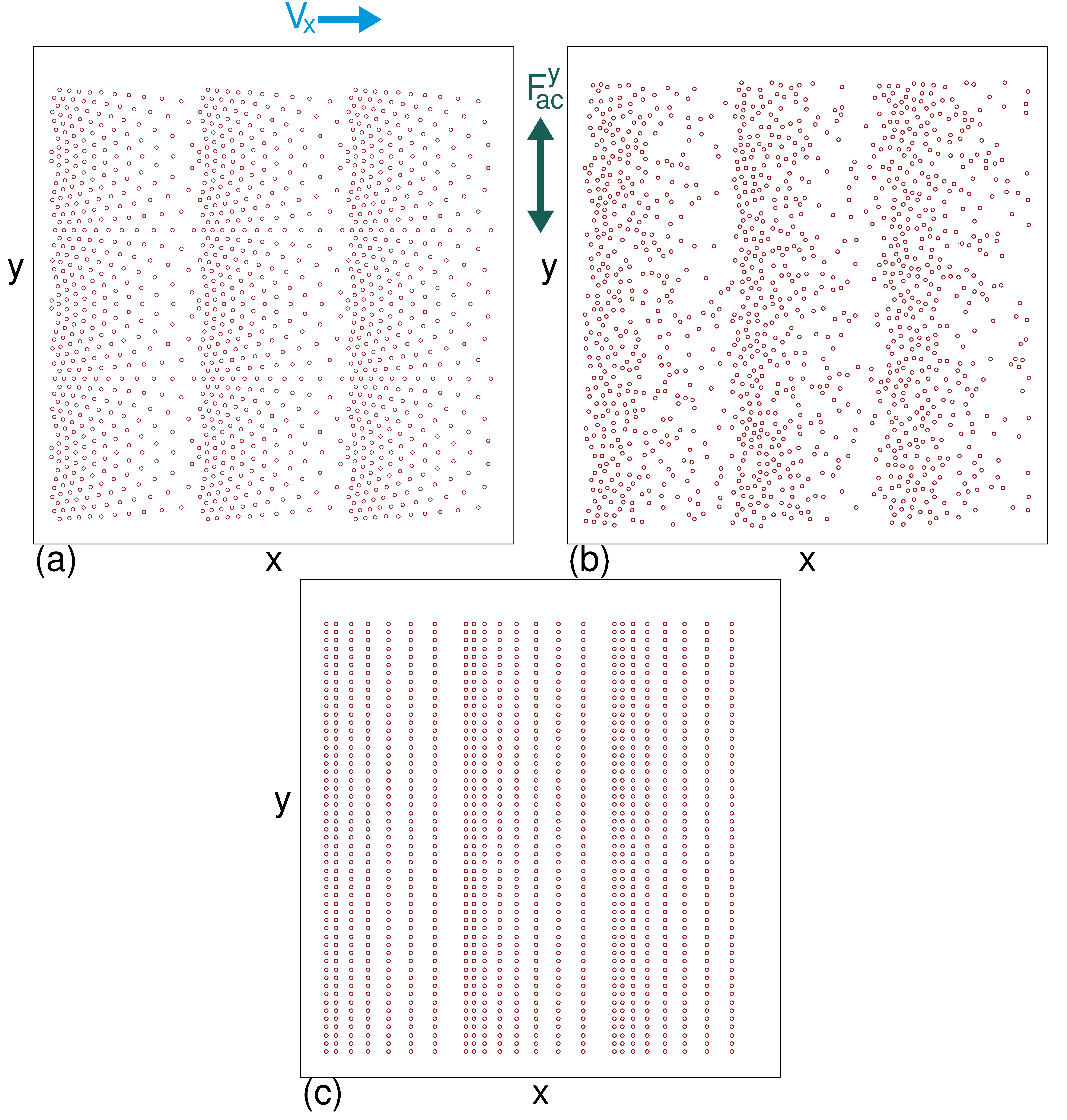}
\caption{
  Circles: the pinning site locations for each geometry.  The substrate asymmetry
  direction is always along the $x$ axis.
  Blue arrow indicates the direction of the ratchet velocity $V_x$.
  (a) Three conformal crystal pinning arrays (ConfG).
  (b) Randomly placed pinning sites with a repeating 1D spatial gradient (RandG).
  (c)  A square pinning array with a series of 1D spatial gradients (SquareG).
  We consider either ac driving along the $y$ direction (dark green arrow)
  or dc driving  along the positive $y$ direction,
  and measure $X_{\rm net}$, the net displacement per vortex in the $x$-direction,
  to quantify the transverse ratchet and geometric ratchet effects.    
}
\label{fig:1}
\end{figure}

Figure~1(a) shows the pinning geometry with three conformal crystals,
each of width $a_{p} = 12\lambda$, placed sequentially along the $x$ direction
of the sample.  We refer to this geometry as ``ConfG.''
Each 
conformal array is generated by performing a conformal transformation of a
uniform triangular lattice in the complex $z$ plane, 
$z = n_{1}b + n_{2}\exp(i\pi/3) b$, where $n_{1}$ and $n_{2}$
are integers and $b$ is the lattice constant.
The transformation that  maps the points from the original lattice to 
the $w$ plane is
\begin{equation}
  w = \frac{\pi}{2\alpha} + \frac{1}{i\alpha}\ln(i\alpha z)
\end{equation}
where $\alpha$ is a parameter. This transformation maps a semi-annular section
covering the region $r_{\rm in}\leq |z| \leq r_{\rm out}$ of the original lattice
to a rectangular region.
We use $\alpha = \pi/36$, an outer annulus radius of $r_{\rm out}=1/\alpha$,
an inner annulus radius of $r_{in} = (1/\alpha)\exp(-\pi/3)$,
and set the lattice constant to
$b = \sqrt{(1 - \exp(-2\pi/3)(\sqrt{3}/\pi)}$.
As shown in Fig.~\ref{fig:1}(a) the easy flow direction is along the
negative $x$ direction. 
We also consider a 
system with a spatial density gradient of randomly placed pinning 
sites, shown in Fig.~\ref{fig:1}(b) and referred to as ``RandG,'' as well as a 
non-conformal gradient array constructed by introducing a one-dimensional density
gradient to a regular square lattice, illustrated in Fig.~\ref{fig:1}(c)
and referred to as ``SquareG.''
  
In previous work examining longitudinal ratchet effects, the ac drive was applied along
the $x$-direction or parallel to the substrate asymmetry direction,
and the net motion was also measured 
along the asymmetry direction \cite{52}.
In the present work  we apply the ac driving
along the $y$-direction or perpendicular to the substrate symmetry direction, as
indicated in Fig.~\ref{fig:1}, but still measure the net flow of vortices along the
$x$ direction, such that a finite dc flow signature indicates the existence of a
transverse ratchet effect.
The ac driving term is ${\bf F}^{ac}_{i} = F^{y}_{ac}\sin(\omega t){\hat {\bf y}}$,
where $F^{y}_{ac}$ is the ac amplitude. 
To measure the ratchet effect we sum the vortex displacements in the direction 
perpendicular to the ac drive
to obtain $X_{\rm net} = N^{-1}_{v}\sum^{N_{v}}_{i=1}(x_{i}(t) - x_{i}(t_{0}))$, 
where $x_{i}(t)$ is the $x$ position of vortex $i$ at time $t$ and
$x_i(t_{0})$ is the initial
position of the vortex when the ac drive is first applied.
We also measure the corresponding $Y_{\rm net}$ using the net displacements along the
$y$ direction, and find that, due to the lack of substrate asymmetry along the
$y$ direction, $Y_{\rm net}=0$ 
for all ac drives that we consider.
We focus on the case with a fixed ac period of 8000 simulation time steps
unless otherwise noted,
and allow the system to run for 50 ac cycles before beginning the measurement
of $X_{\rm net}$ 
to avoid any transient effects, as was done in previous studies of longitudinal 
ratchet effects \cite{52}.  
We also consider the case $F^y_{ac} = 0$ in which we
apply a finite dc force ${\bf F}^{dc}_{i} = F^{y}_{dc}{\hat {\bf y}}$
along the positive $y$-direction,
and examine $X_{\rm net}$ after the equivalent of 50 ac cycles of time has passed
to quantify  the geometric ratchet effect. 
Finally, we consider the effect of adding thermal fluctuations using
${\bf F}^{T}_{i}$, which has the properties
$\langle F^{T}_{i}(t)\rangle = 0$ and $\langle F^{T}_{i}(t)F^{T}_{j}(t^{\prime})
\rangle =  2\eta k_{B}T\delta_{ij}\delta(t-t^{\prime})$,
where $k_B$ is the Boltzmann constant.
Unless otherwise noted, we set $F^T=0$.

\section{Transverse ac Ratchet Effect}

\begin{figure}
\includegraphics[width=3.5in]{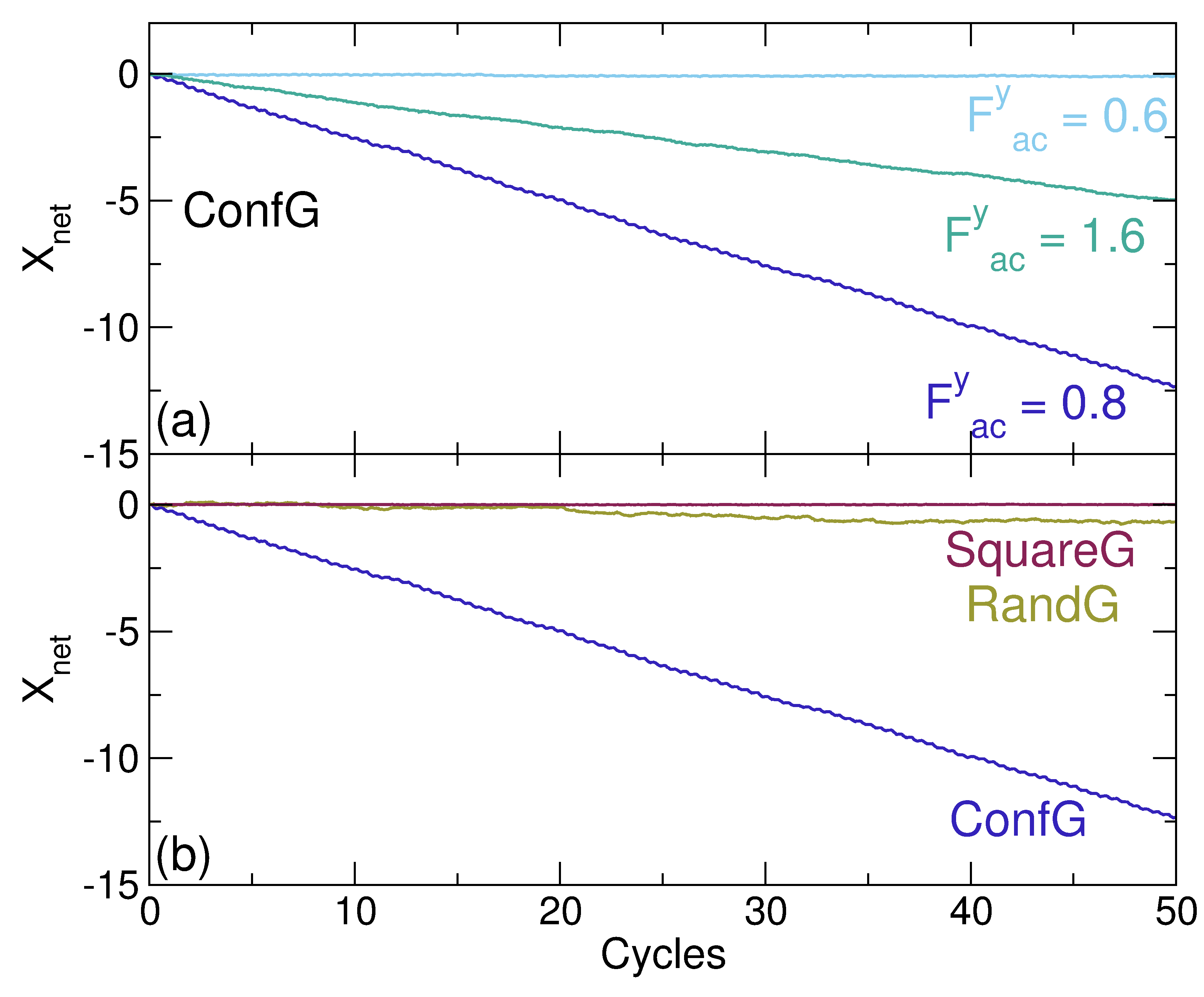}
\caption{$X_{\rm net}$, the average displacement per vortex in
  the longitudinal $x$-direction, vs ac cycle number
  for an ac drive $F^y_{ac}$ applied in the transverse or $y$ direction in samples with
  $B/B_{\phi}=1.0$ and  $F_p=1.0$.
  (a) The ConfG conformal array in  Fig.~\ref{fig:1}(a)
  at $F^{y}_{ac} = 0.6$ (upper light blue curve) where there is no ratchet effect,
  at $F^{y}_{ac} = 0.8$ (lower dark blue cuve) where there is a strong ratchet effect, and
  $F^{y}_{ac} = 1.6$ (middle green curve) where there is a finite but reduced ratchet effect.
  (b) At $F^{y}_{ac}=0.8$, 
the SquareG square gradient array from Fig.~\ref{fig:1}(c) (upper red curve)
  has no ratchet effect,
  the RandG random gradient array from Fig.~\ref{fig:1}(b) (middle brown curve) has
  a weak ratchet effect,
  and the ConfG array (lower dark blue curve) has a strong ratchet effect.  The conformal array
  generates a ratchet effect that
  is approximately
  20 times more effective than that of the RandG array. In all cases $Y_{\rm net} = 0.0$.     
}
\label{fig:2}
\end{figure}

In Fig.~\ref{fig:2}(a) we plot
$X_{\rm net}$, the net displacement per vortex in the $x$-direction, verus time in
ac drive cycles for the ConfG
conformal array in Fig.~\ref{fig:1}(a) at $B/B_{\phi} = 1.0$ and $F_{p} = 1.0$.
For $F^y_{ac} = 0.6$,  $X_{\rm net} = 0$ indicating that there is
no transverse ratchet effect.
At $F^{y}_{ac} = 0.8$ a finite transverse ratchet effect emerges and the vortices
each move an average of $12.5\lambda$ in the negative $x$ direction during 50 ac drive
cycles.
The vortices translate along the easy flow direction of the substrate,
and the displacement 
increases linearly with time indicating that transient effects are not present.
For $F^{y}_{ac} = 1.6$, ratcheting still occurs but the effect is reduced,
with the vortices moving an average of $5.0\lambda$ in the negative
$x$ direction during 50  ac drive cycles.

\begin{figure}
\includegraphics[width=3.5in]{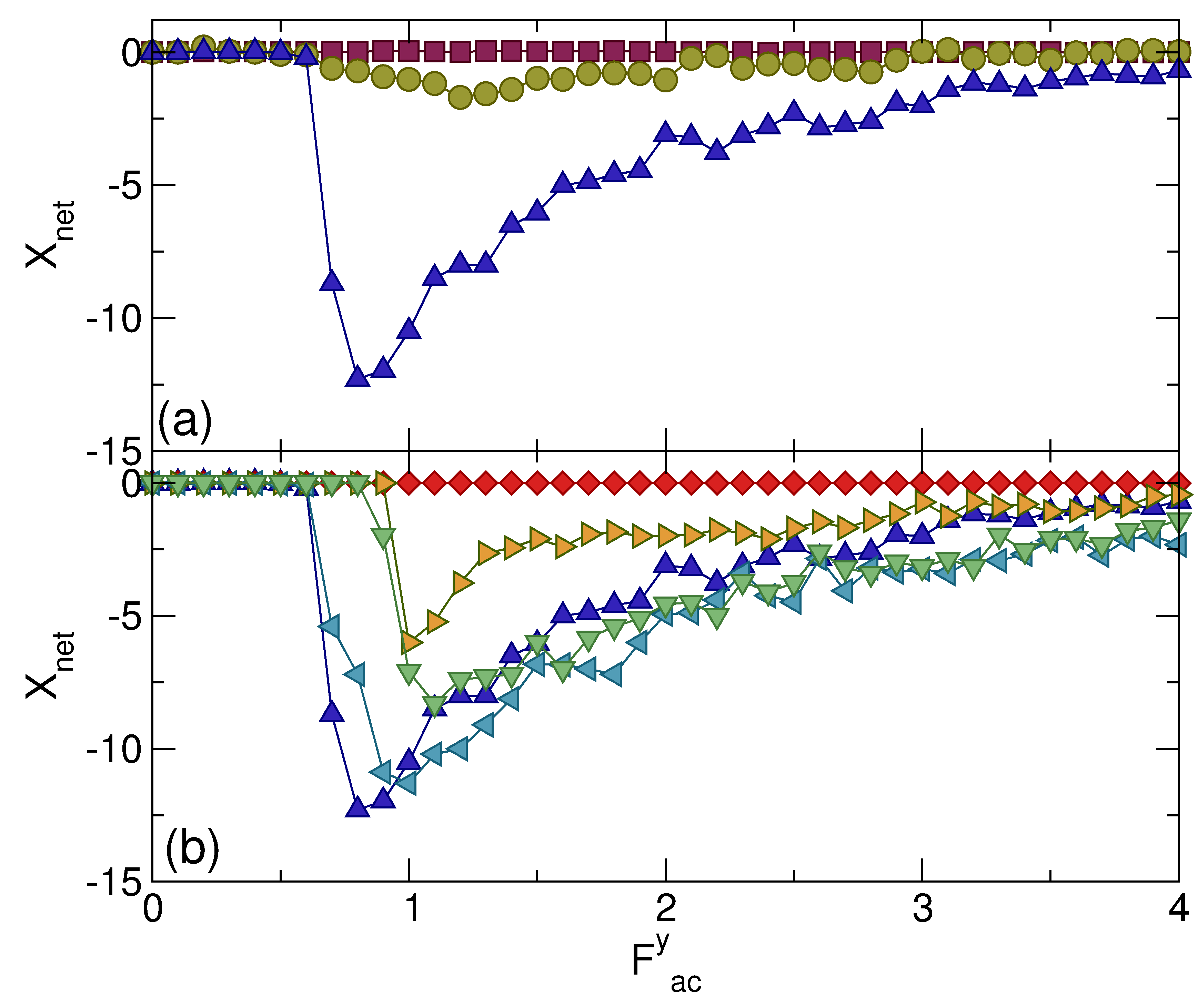}
\caption{(a) $X_{\rm net}$ after 50 ac drive cycles versus $F^y_{ac}$
  for samples with $B/B_{\phi} = 1.0$ and $F_{p} = 1.0$.
 Red squares: SquareG array; brown circles: RandG array; dark blue triangles: ConfG array.
  The ConfG array produces the strongest ratchet effect, while the ratchet effect is
  absent in the SquareG array.
(b) $X_{\rm net}$ after 50 ac drive cycles vs $F^{y}_{ac}$ for a ConfG array
with $F_{p} = 1.0$
at $B/B_{\phi} = 1.0$ (dark blue up triangles), $0.5$ (light blue left triangles),
$0.2$ (green down triangles),
$0.1$ (orange right triangles), and $0.05$ (red diamonds), showing that collective effects
are important for the transverse ratchet motion.
}
\label{fig:3}
\end{figure}

Figure~\ref{fig:2}(b) shows $X_{\rm net}$ at $B/B_{\phi}=1.0$, $F_p=1.0$, and
$F^{y}_{ac} = 0.8$ for the different array types.
In the SquareG square gradient array
from Fig.~\ref{fig:1}(c), $X_{\rm net} = 0$ after 50 ac cycles,
while the RandG random gradient array from Fig.~\ref{fig:1}(b) exhibits
ratcheting that is 20 times less effective than in the ConfG array, which is also shown
for comparison.
To better quantify the ratchet 
as a function of $F^{y}_{ac}$, in Fig.~\ref{fig:3}(a) we plot the value of
$X_{\rm net}$ at the end of $50$ ac drive cycles for ConfG, RandG, and SquareG arrays
at $F_{p} = 1.0$ and $B/B_{\phi}  = 1.0$. 
In the ConfG array,
$X_{\rm net} = 0.0$ when $F^{y}_{ac} < 0.6$, and the ratchet reaches
its maximum efficiency 
near $F^{y}_{ac}  = 0.8$, after which $X_{\rm net}$ gradually approaches zero
with increasing $F^y_{ac}$.
The RandG array shows similar behavior, but has a much weaker overall ratchet
effect and exhibits an efficiency maximum at
$F^{y}_{ac} = 1.2$.
In the SquareG array,
$X_{\rm net} = 0$ for all values
of $F^{y}_{ac}$. 

To explore the role of vortex-vortex interactions in the ratchet effect, in 
Fig.~\ref{fig:3}(b) we plot the value of
$X_{\rm net}$ after 50 ac drive cycles versus $F^{y}_{ac}$
for the ConfG system in Fig.~\ref{fig:3}(a) at
$B/B_{\phi} = 1.0$, $0.5$, $0.2$, $0.1$, and $0.05$.
The overall effectiveness of the ratchet decreases with
decreasing $B/B_{\phi}$, and for $B/B_{\phi} < 0.1$ the ratchet effect 
is absent.
Whenever $F^{y}_{ac}>F_p$, there is a portion of the ac cycle during which all of
the vortices are depinned and moving, so the loss of the ratchet effect at low
$B/B_{\phi}$ is not caused by the vortices becoming pinned when their density is
small.  Instead, this result indicates that
collective vortex-vortex 
interactions are important for the transverse ratchet to occur.

\begin{figure}
\includegraphics[width=3.5in]{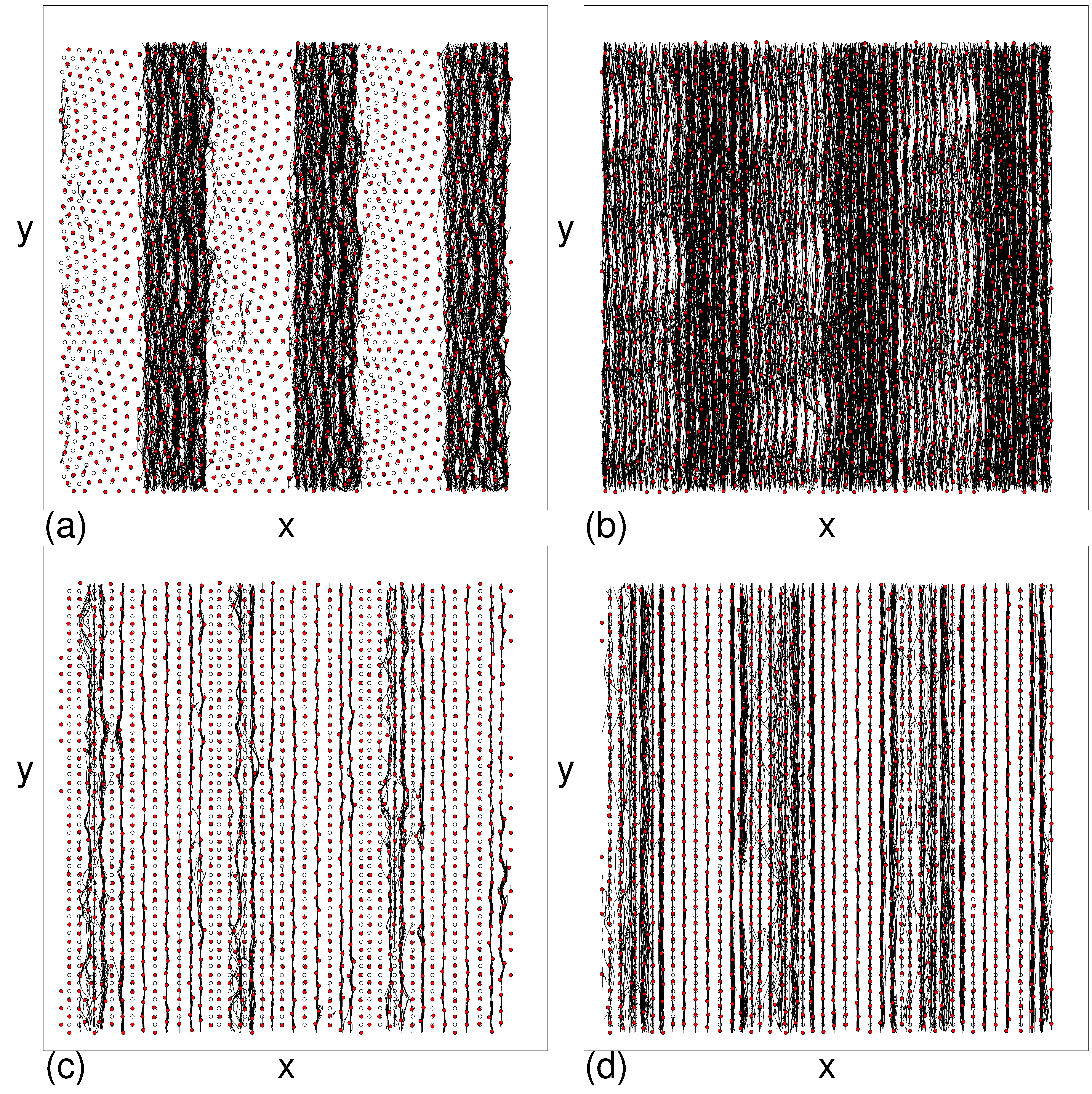}
\caption{The pinning site locations (open circles), instantaneous vortex positions
  (red dots), and vortex trajectories (lines)
  for samples with
  $B/B_{\phi} = 1.0$ and $F_{p} = 1.0$.
  (a) In the ConfG array at $F^{y}_{ac} = 0.5$ there is no ratchet effect.
  (b) The ConfG array at $F^{y}_{ac} = 0.8$ shows a ratchet effect.
  (c) There is no ratchet effect in the SquareG array at $F^y_{ac}=0.5$.
  (d) There is also no ratchet effect in the SquareG array at $F^y_{ac}=0.8$.
  The vortex motion is more one-dimensional along the $y$-direction in the SquareG array
  than in the ConfG array.
}
\label{fig:4}
\end{figure}

In Fig.~\ref{fig:4}(a) we plot the instantaneous vortex positions, vortex trajectories,
and pinning site locations for the ConfG array in Fig.~\ref{fig:3}(a) 
at $F^{y}_{ac} = 0.5$ where there is no ratcheting.
Here the vortex motion is confined to the low pinning density regions of the sample.
For lower values of  $F^{y}_{ac}$, the width of the regions of moving vortices
decreases.
For $F^{y}_{ac} > 0.6$,
all the vortices
are able to move,
as shown in Fig.~\ref{fig:4}(b) 
for $F^{y}_{ac} = 0.8$ where finite ratcheting in the
negative $x$-direction occurs.
The vortices do not move strictly along the $y$ or driving direction, but follow winding
trajectories that introduce strong $x$-direction velocity fluctuations.
In a SquareG array with the same parameters,  no ratcheting occurs.
Figure~\ref{fig:4}(c) shows that 
at $F^{y}_{ac} = 0.5$ in the SquareG array,
the vortex trajectories are strongly one-dimensional and are oriented along the $y$
direction with few or no fluctuations along the $x$ direction.
In Fig.~\ref{fig:4}(d), at $F^{y}_{ac} = 0.8$ for the SquareG array all the
vortices can participate in the flow during some portion of the ac cycle, 
but again the  motion follows nearly straight trajectories
along the $y$-direction and there is no ratchet effect.
As $F^{y}_ {ac}$ increases above $F^{ac}_{y}=0.8$, the
$x$ direction meandering of the vortex trajectories in the ConfG array
is progressively reduced, and this
coincides with the drop in ratchet efficiency shown in
Fig.~\ref{fig:3}(a).  The vortex trajectories for the RandG array are similar in appearance
to those shown for the ConfG array.

\begin{figure}
\includegraphics[width=3.5in]{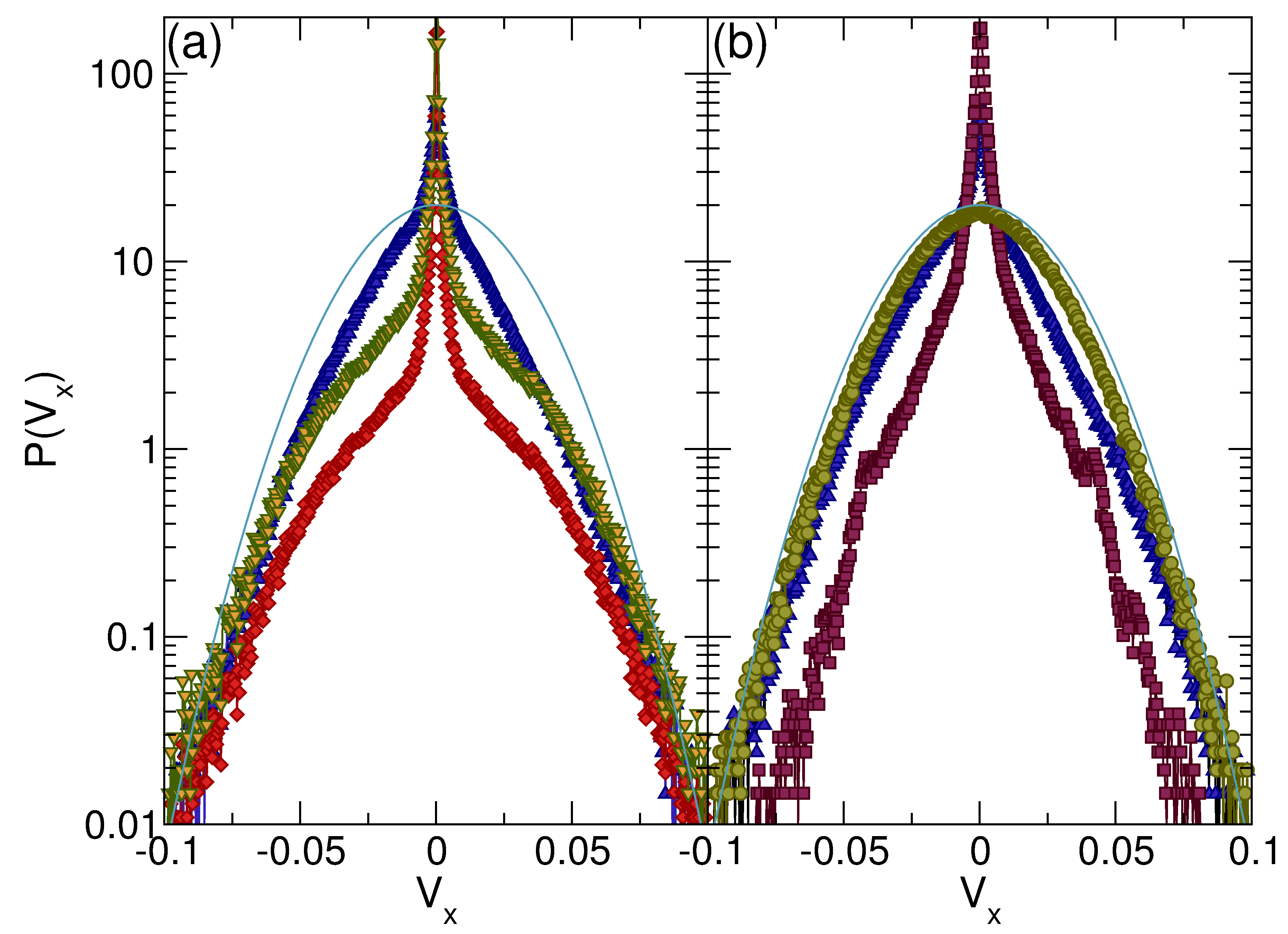}
\caption{$P(V_x)$, the distribution of individual vortex velocities obtained over
  50 ac driving cycles, for the system in Fig.~\ref{fig:3} at
$B/B_{\phi} = 1.0$ and $F_{p} = 1.0$.
  The solid light blue line is a plot of a
Gaussian curve with $P(V_x)=V_{0}\exp(-\alpha x^2)$. 
  (a) The ConfG array at
  $F^{y}_{ac} = 0.5$ (red diamonds) where there is no ratchet effect,
  $F^{y}_{ac}=0.8$ (dark blue up triangles) where there is a strong ratchet effect,
  and $F^y_{ac}=3.5$ (orange down triangles) where there is a weak ratchet
  effect.
(b) $P(V_x)$ at  $F^{y}_{ac} =0.8$ for
the SquareG array (red squares) where there is no ratchet effect,
the ConfG array (dark blue triangles) where there is a strong ratchet effect,
and the RandG array (brown circles) where there is a weak ratchet effect.
The fluctuations for the RandG array are
nearly Gaussian.
}
\label{fig:5}
\end{figure}

The transverse ratchet in the ConfG and RandG arrays can be understood as 
a realization of a noise correlation
ratchet where the correlated noise is
generated by the plastic flow of the vortices.
To clarify this, we examine the  
$x$-component velocity distributions $P(V_{x})$ of the individual vortices over a fixed
time of 50 ac drive cycles.
In Fig.~\ref{fig:5}(a) we plot 
$P(V_{x})$ for the ConfG system from Fig.~\ref{fig:3} with
$B/B_{\phi} = 1.0$ and $F_{p} = 1.0$.  
At $F^{y}_{ac} = 0.8$, where there is a strong ratchet effect,
$P(V_{x})$ differs significantly from the
simple Gaussian form
$P(V_x)=V_{0}\exp{-\alpha x^2}$,  
which is plotted
as a smooth solid line.
For $F^{y}_{ac} = 0.5$ where there is no ratcheting,
there is a strong peak in $P(V_x)$ at $V_{x} = 0$ due to the pinned vortices,
and the magnitude of the $x$ velocity fluctuations are
significantly reduced compared to the $F^{y}_{ac} = 0.8$ case.
For $F^{y}_{ac} = 3.5$ where the
ratchet effect is present but weak, as shown in Fig.~\ref{fig:3}(a),
the width of $P(V_{x})$ is smaller than at the optimal ac drive of
$F^{y}_{ac}=0.8$.

In Fig.~\ref{fig:5}(b) we plot $P(V_x)$ for the ConfG, SquareG, and RandG arrays
from Fig.~\ref{fig:3}(a)
at $F^{y}_{ac} = 0.8$.
The width of $P(V_x)$ is much narrower for the SquareG
array than for the ConfG and RandG arrays due to the strongly 1D nature
of the vortex trajectories in the SquareG array, as shown in Fig.~\ref{fig:4}(d).
In the RandG array, $P(V_x)$ for $V_x<-0.01$ is nearly identical to that of the
ConfG array; however, close to $V_x=0$ the RandG array lacks the pinned vortex
peak found in the ConfG array and instead
maintains a Gaussian
form of $P(V_x)=V_0 \exp(-\alpha x^2)$.
For $V_x>0.01$, the ConfG array has a reduction in $P(V_x)$ compared to the RandG
array; this asymmetry in $P(V_x)$ is discussed further in Section IV.B.
The Gaussian nature of the
dynamically generated fluctuations in the RandG array produces weak time correlations
in the velocity, leading to a ratchet effect that is weaker than that found in the ConfG
array.  In the SquareG array the velocity is strongly peaked at a single value, meaning
that there are insufficient velocity fluctuations to generate a ratchet effect.

\subsection{Transverse Ratchet Reversals}

\begin{figure}
\includegraphics[width=3.5in]{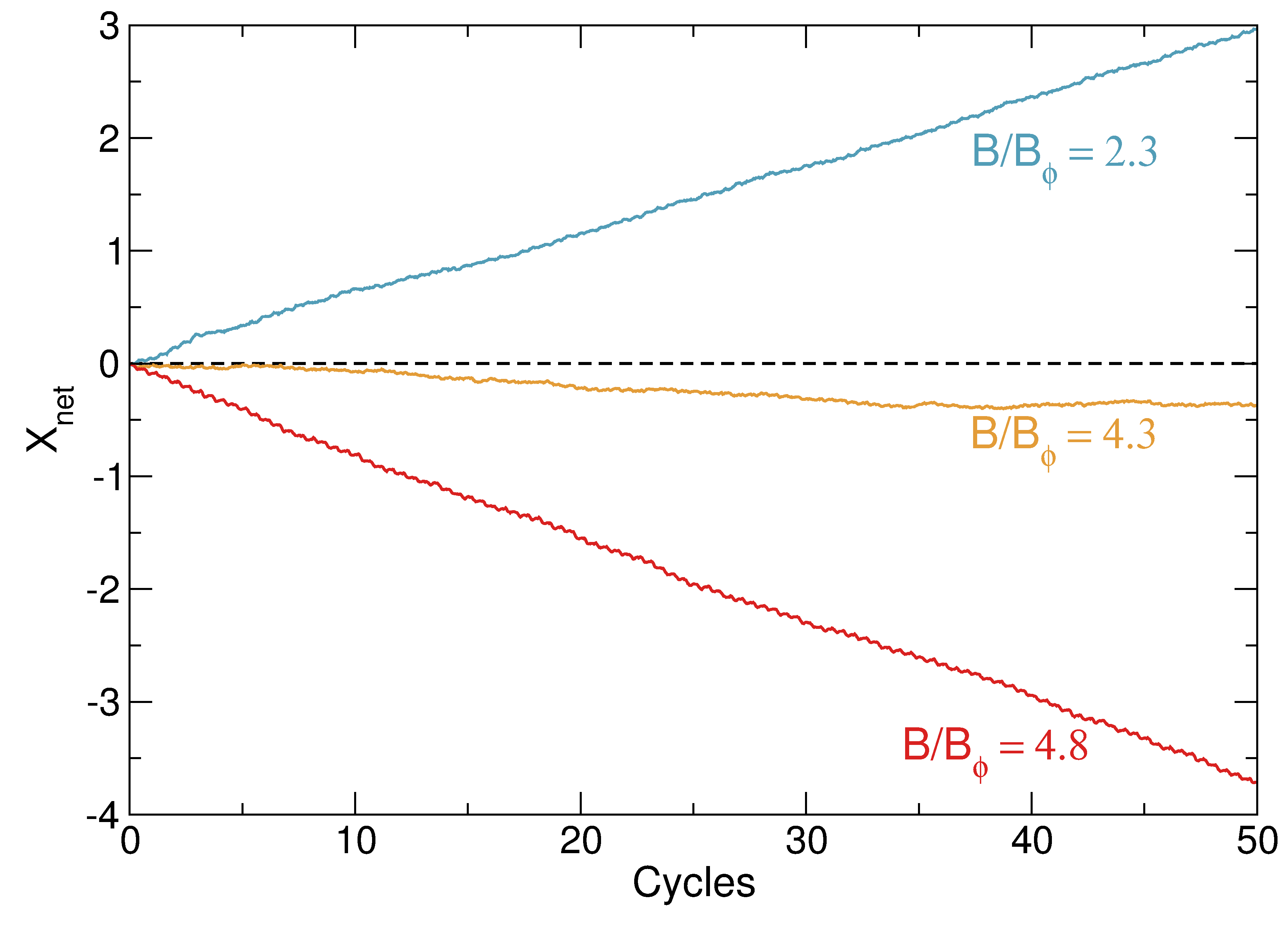}
\caption{ $X_{\rm net}$ vs
  ac drive cycle number for the
  ConfG array at $F_{p} = 2.0$ and $F^{y}_{ac} = 0.7$.
  Upper blue curve: At $B/B_{\phi} = 2.3$ there is  a ratchet effect
  in the positive $x$ or hard direction of the
  substrate asymmetry, referred to as a reversed transverse ratchet effect.
  Middle orange curve: At $B/B_{\phi} = 4.3$ there is a weak
  negative $x$ or easy substrate asymmetry direction ratchet effect,
  referred to as a normal transverse ratchet effect.
  Lower red curve: At $B/B_{\phi} = 4.8$ there
  is a much stronger normal transverse ratchet effect.   
}
\label{fig:6}
\end{figure}

\begin{figure}
\includegraphics[width=3.5in]{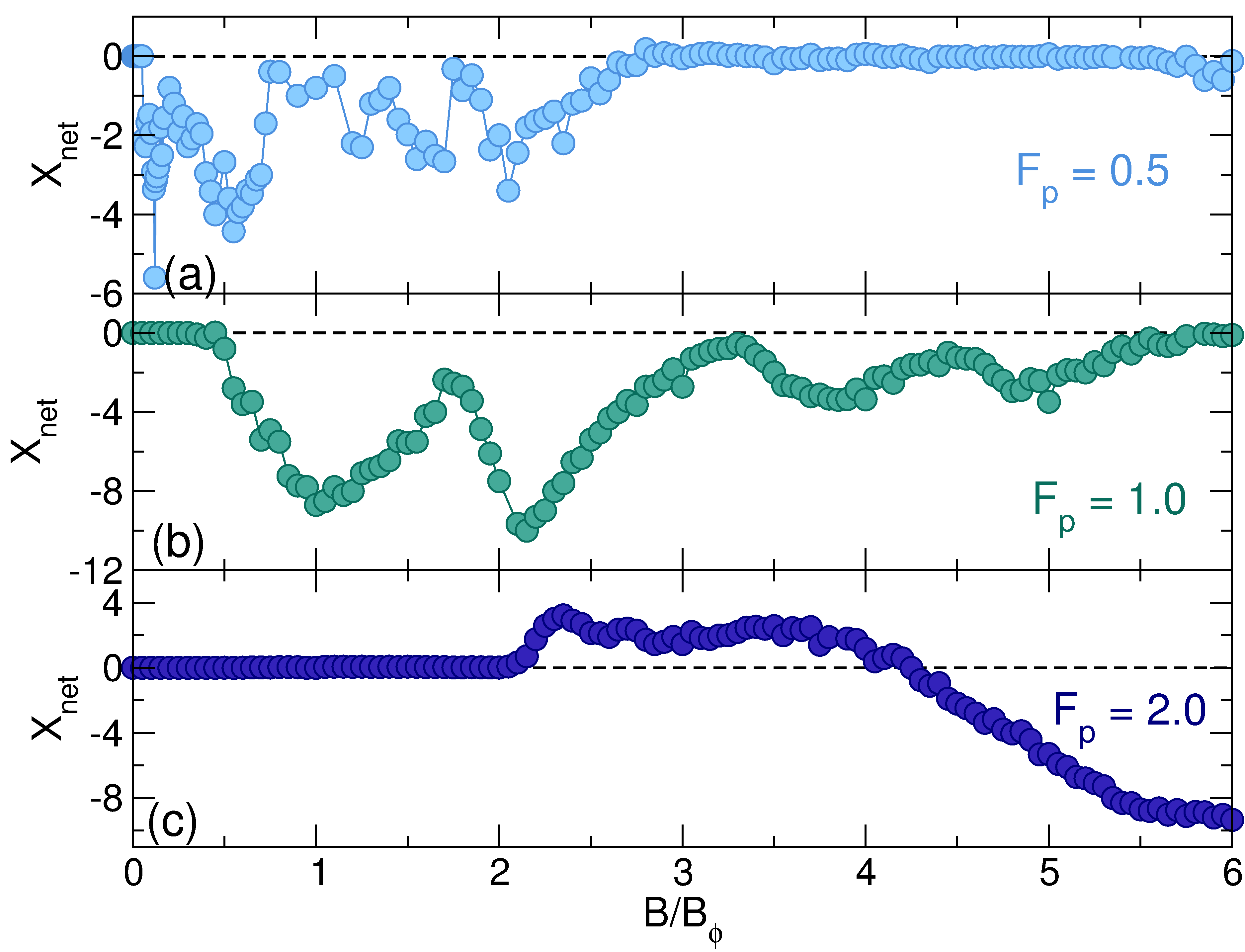}
\caption{$X_{\rm net}$ after 50
  ac drive cycles vs $B/B_{\phi}$ in a ConfG array with
  $F^{y}_{ac} = 0.7$.
  (a) At $F_{p} = 0.5$ there is a normal transverse ratchet effect.
  (b) At $F_{p} = 1.0$ there are large variations in the magnitude of the
  transverse ratchet effect.  
  (c) At $F_{p} = 2.0$ there is a reversal in the vortex ratchet effect
  from a reversed transverse ratchet effect
  with motion in the positive $x$ direction for $2<B/B_{\phi}<4.2$ to
  a normal transverse ratchet effect with motion in the negative
  $x$ direction for $B/B_{\phi}>4.2$.
}
\label{fig:7}
\end{figure}

It is also possible to realize a reversal 
of the transverse ratchet effect where the net flow of vortices
is in the positive $x$ or hard
direction of the substrate asymmetry.
Such motion is termed a reversed ratchet effect, and it is marked by net flow in the
positive $x$ direction due to the orientation of our ratchet potential.
In Fig.~\ref{fig:6} we plot $X_{\rm net}$ as a function of time
for a ConfG array with
$F_{p} = 2.0$ and $F^{y}_{ac} = 0.7$.
At $B/B_{\phi} = 2.3$, the vortices are translating in the positive $x$
direction, corresponding to a reversed ratchet effect,
while at  $B/B_{\phi}  = 4.3$ there is a weak
normal ratchet effect in the negative $x$ direction
and at $B/B_{\phi} = 4.8$ there is a stronger normal ratchet effect in the negative
$x$ direction, showing that 
a ratchet reversal occurs as a function of vortex density.   
In Fig.~\ref{fig:7} we show 
$X_{\rm net}$ versus $B/B_{\phi}$ for the ConfG array at fixed 
$F^{y}_{ac} = 0.7$ and varied $F_{p}$.
For $F_p=0.5$ in Fig.~\ref{fig:7}(a), there is a normal ratchet effect in the
negative $x$ direction with a magnitude that is largest for $B/B_{\phi}<1.0$.
There are local maxima in the ratchet effectiveness
near $B/B_{\phi} = 0.5$ and 
$B/B_{\phi} = 2.0$, and the ratchet effect
disappears for $B/B_{\phi} > 3.0$ since the 
pinning is weak enough that 
the vortices start to form a uniform triangular lattice at the higher fields.
In Fig.~\ref{fig:7}(b) at $F_{p} = 1.0$,  there is a normal ratchet 
effect that is suppressed for $B/B_{\phi} < 0.5$ where the vortices are unable to
depin.
The ratchet effect is larger in magnitude than for the $F_{p} = 0.5$ case,  
and the maximum ratchet effectiveness occurs just above $B/B_{\phi} = 2.0$. 
The strong variations in the ratchet effect as a function of field
reflect the occurrence of partial commensuration effects, with the most pronounced
ratchet motion appearing near
$B/B_{\phi} = 1.0$, 2.0, 4.0, and $5.0$.   
Interestingly, there is no commensuration effect near $B/B_{\phi} = 3.0$;
this is in contrast with
previous studies of vortex ordering in uniform triangular lattices, where
ordered triangular vortex lattices
associated with  peaks in the critical depinning force occur
at matching fields of $B/B_{\phi} = 1.0$, 3.0, and $4.0$,  
while there is much weaker matching at $B/B_{\phi}  = 2.0$
and $5.0$ when ordered but nontriangular vortex lattices form \cite{58}.
The commensurability effects may also be different for square conformal arrays, as
studies of uniform square pinning arrays show that different kinds of vortex configurations
are stable at different integer matching fields \cite{58,New1,New2}.
Due to the gradient in the ConfG array,
commensuration conditions can occur in only part of the sample
at a time, as illustrated in previous simulations,
so that the matching effects are not centered at integer ratios of $B/B_{\phi}$ \cite{51}.
In Fig.~\ref{fig:7}(c) we plot $X_{\rm net}$ versus $B/B_{\phi}$ for
the same ConfG system at $F_{p} = 2.0$ that is highlighted in
Fig.~\ref{fig:6}.  The 
ratchet effect is absent for $B/B_{\phi} < 2.0$, while for
$2.0 < B/B_{\phi} < 4.0$ the vortices
exhibit a reversed ratchet effect and flow in the positive $x$ direction.
For $B/B_{\phi}>4.0$ the vortices flow in the negative $x$ direction to produce
a normal ratchet effect.

The switch from a reversed to a normal ratchet effect
occurs due to changes in the $x$-direction fluctuations of the vortices
moving
along the pinning gradient.
When $F_p$ is weak,
vortex motion occurs across the entire pinning gradient and
the largest transverse fluctuations of the flowing vortices
take place in the least densely  pinned
portions of the sample, 
while vortices spend more time pinned
in the most densely pinned regions, producing smaller
transverse velocity fluctuations as
shown in Fig.~\ref{fig:4}(a).
In analogy 
with the thermophoretic effect, in which particles preferentially drift
from hotter to colder portions of a sample  \cite{59},
the vortices tend to drift from the low pinning density regions to the high pinning density
regions, so that within an individual substrate ratchet plaquette, the
vortices move in the negative $x$ direction.
For the normal transverse ratchet effect, the pinning establishes a vortex density
gradient that is maximum on the high pinning density side of each substrate ratchet
plaquette, and this vortex density gradient breaks an additional symmetry for
the fluctuation-induced vortex drift, preventing vortices from moving in the
positive $x$ direction in order to
pass directly from
the lowest pinning density strong velocity fluctuation region
to the highest pinning density small velocity fluctuation
region.
When $F_p$ is strong, the flow in the regions with low pinning density
primarily consists of interstitial vortices that are not trapped in pinning sites
moving between occupied pinning sites.  The resulting winding flow creates 
smooth velocity fluctuations in the $x$ direction.
In the regions with high pinning density,
the small spacing between pinned vortices forces the
interstitial vortices to approach the pinned vortices much more closely,
and the resulting vortex-vortex interaction forces depin the pinned vortices,
generating enhanced fluctuations in $V_x$ in the high pinning density region.
As a result, the effective temperature gradient induced by the velocity fluctuations
is reversed compared to the case of low $F_p$, leading to a reversal of the ratchet
flow direction.

\begin{figure}
\includegraphics[width=3.5in]{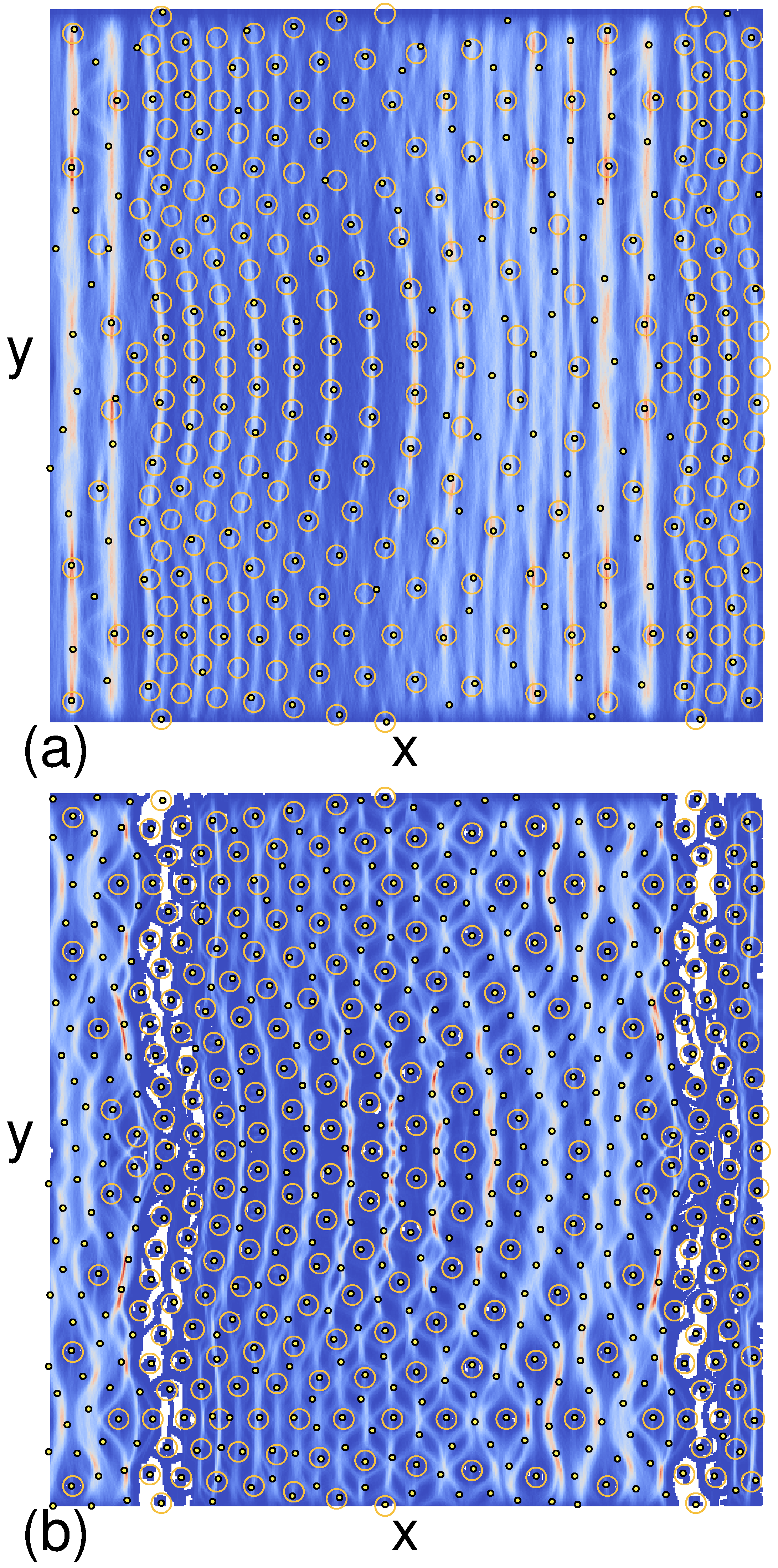}
\caption{The pinning site locations (orange circles), 
  vortex locations (red circles), and heightfield of accumulated vortex trajectories
  (blue: few trails; red: many trails)
  in a $10\lambda \times 10\lambda$ portion of the
  ConfG sample.
  (a) At $F_p=1.0$, $B/B_{\phi}=1.0$, and $F^{y}_{ac}=0.75$, there
  is a normal ratchet effect in
  the negative $x$ direction.  Vortices pass through the pinning sites and the largest
  $x$-direction fluctuations of the trajectories occur in the low pinning density regions.
  (b) At $F_p=2.0$, $B/B_{\phi}=2.3$, and $F^{y}_{ac}=0.75$, 
  there is a reversed ratchet effect
  in the positive $x$-direction.   Interstitial vortices flow around the
  pinned vortices.  The blank white regions in the densest portion of the pinning array are
  locations through which vortices never flow.
  The $x$-direction fluctuations of the trajectories
  are enhanced in the high pinning density portion of the sample.
}
\label{fig:8}
\end{figure}

In Fig.~\ref{fig:8} we show a heightfield of the vortex trajectories in a portion of the
ConfG sample, obtained by rasterizing the vortex trajectories onto a fine grid over the
course of 50 ac drive cycles and measuring the total number of trails that pass through
each grid point.
At $F_p=1.0$, $B/B_{\phi}=1.0$, and $F^{y}_{ac}=0.75$ in Fig.~\ref{fig:8}(a),
the ratchet effect is in the normal negative $x$-direction.  Vortices flow through the
pinning sites, and the vortex trajectories fluctuate the most in the low pinning density
portion of the sample.  In the high pinning density area,
the channeling of the vortices through
successive pinning sites suppresses velocity fluctuations transverse to the driving
direction.
At $F_{p} = 1.0$, $B/B_{\phi} = 2.3$, and $F^{y}_{ac} = 0.75$,
the ratchet effect is reversed and the vortices translate in the positive $x$ direction.
In the highest pinning density portion of the sample,
vortex motion occurs via a combination of purely interstitial vortex flow in the middle of
the illustrated region and hopping of vortices from one pinning site to the next at the top
and bottom of the illustrated region, producing enhanced
velocity fluctuations in the $x$ direction.
In the regions with low pinning density, the vortices at the pinning sites remain pinned
most of the time and 
the vortex motion consists almost entirely of smooth interstitial flow with reduced
velocity fluctuations in the $x$-direction.
As $F^{y}_{ac}$ is increased,  the vortices in the low pinning density region begin to
depin more frequently, while the flow in the high pinning density region shifts from
partially interstitial to channelling along the pins, shifting the relative magnitude of the
$x$-velocity fluctuations so that it is highest in the low pinning density region, and
switching the ratchet effect back to the normal negative $x$ direction.
In Fig.~\ref{fig:7}(c), where $F^{y}_{ac}$ is held fixed
at $F^{y}_{ac}=0.7$ as $B/B_{\phi}$ is increased,
more vortices occupy the low pinning density regions of the sample as $B/B_{\phi}$
becomes larger, and the increased strength of the vortex-vortex interactions causes
the vortices located at the pinning sites to depin.  As a result, the magnitude of the
$x$-velocity fluctuations in the low pinning density regions
increases as the magnetic field increases, leading to the reversal of the ratchet effect.

\begin{figure}
\includegraphics[width=3.5in]{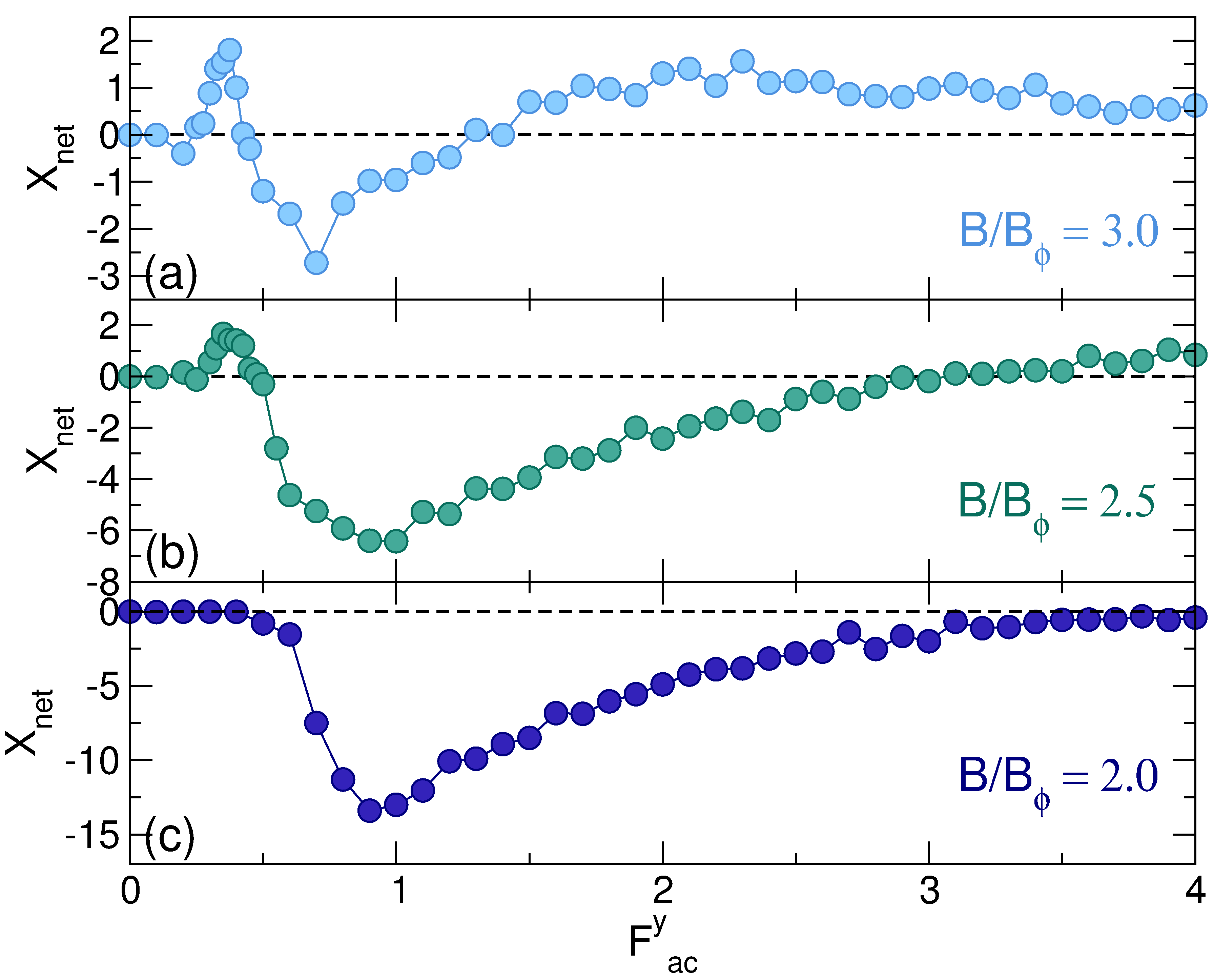}
\caption{$X_{\rm net}$ after 50 ac drive cycles
  vs $F^{y}_{ac}$ for the ConfG array at $F_{p} = 1.0$.
  (a) At $B/B_{\phi} = 3.0$ there are two ratchet reversals.
  (b) $B/B_{\phi} = 2.5$.
  (c) At $B/B_{\phi} = 2.0$ the ratchet effect is always in the normal negative $x$ direction.
}
\label{fig:9}
\end{figure}

The ratchet reversals can also occur at a 
fixed field when the ac driving amplitude is varied, as shown in 
Fig.~\ref{fig:9}(a) where
we plot $X_{\rm net}$ versus $F^{y}_{ac}$ for a ConfG array
with $F_{p} = 1.0$ at $B/B_{\phi} = 3.0$.
There is no ratchet effect for $F^{y}_{ac} < 0.25$.  A reversed ratchet effect
with vortex flow 
in the positive $x$ direction occurs
for $0.25 < F^{y}_{ac} < 0.425$, followed by
a region of normal ratchet motion in the negative $x$ direction for
$0.425 < F^{y}_{ac} < 1.5$.   The ratchet flow is in the reversed positive $x$
direction again for $F^{y}_{ac} > 1.5$. 
In the range $0.25 <F^{y}_{ac} < 0.425$,  the vortex motion
in the high pinning density regions of the sample occurs through a mixed
interstitial and channelling flow of the type illustrated in Fig.~\ref{fig:8}(b),
while for $0.425 < F^y_{ac}<1.5$,
the ac drive is large enough to depin all the vortices, producing
strongly disordered flow throughout the  sample and resulting in a normal negative
$x$ ratchet effect.

\begin{figure}
\includegraphics[width=3.5in]{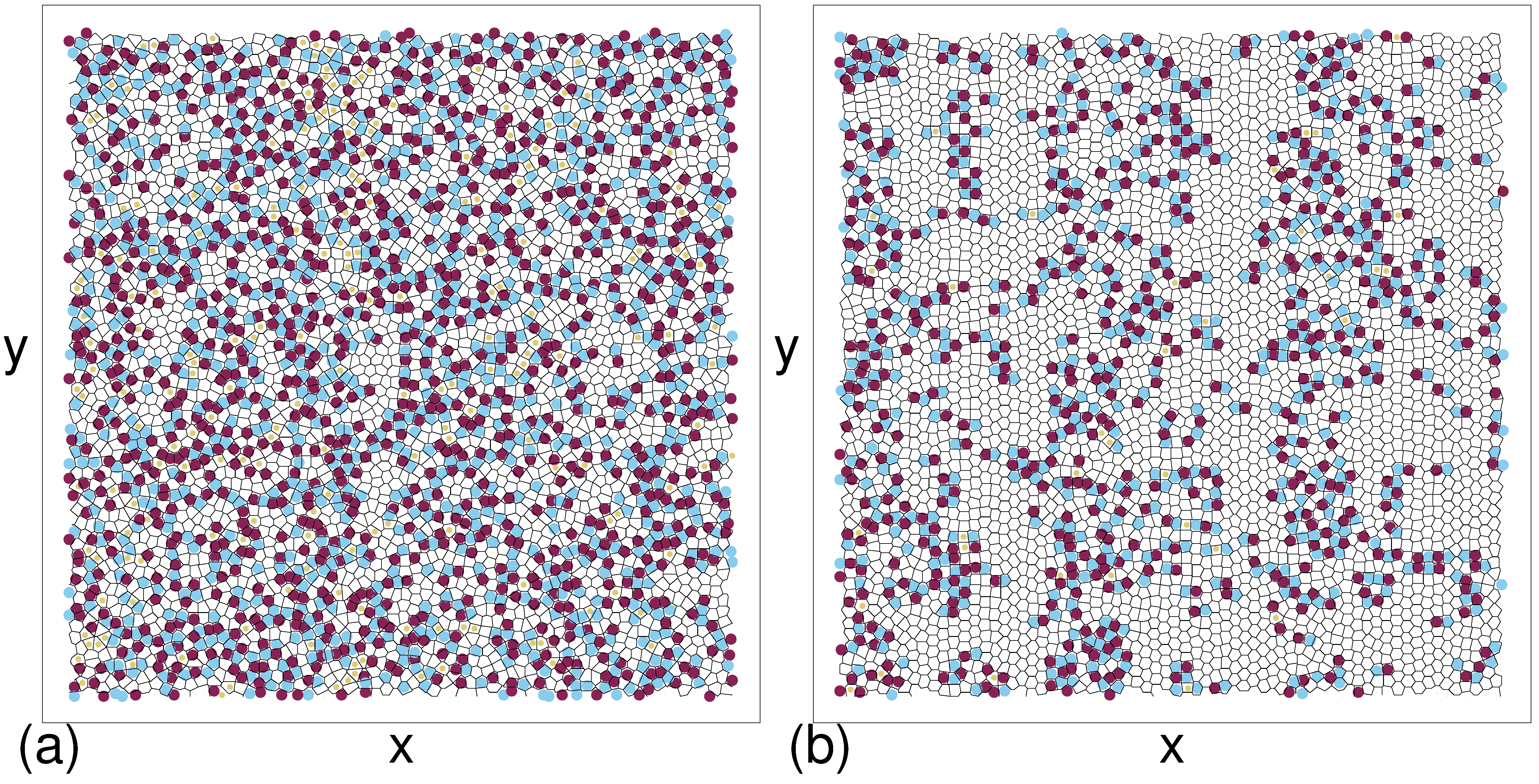}
\caption{The Voronoi construction of the vortex positions
  in the entire sample
  for the system in Fig.~\ref{fig:9}(a)
  showing vortex coordination
  number $z_i$.   White: $z_i=6$; dark red: $z_i=5$; light blue: $z_i=7$; yellow: $z_i=8$.
(a) At $F^{y}_{ac} = 0.7$, the ratchet is in the normal negative $x$ direction.
  (b) At $F^{y}_{ac} = 2.3$ the sample is in the second region of
positive $x$ reversed ratchet motion.  The defect density is largest in the highest pinning
  density portion of the sample, while the low pinning density regions
  are associated with bands of
  ordered vortex lattice.
}
\label{fig:10}
\end{figure}

The mechanism responsible for the appearance of the
second reversed, positive $x$ ratcheting region
for $F^{y}_{ac} > 1.5$ 
differs from that found at lower ac drive.  
At high ac drives, over a portion of the drive cycle the driving current is large enough to
induce dynamical ordering of the vortices in the low pinning density portions of the
sample.  This lowers the $x$ velocity fluctuations in the low pinning density portions of the
sample compared to other areas of the sample where the vortex lattice remains more
disordered.
As a result, the effective shaking temperature is highest in the high pinning density
region of the sample
even though all the vortices are flowing. 
Previous simulations and experiments on dc driven vortices in samples
with uniform pinning density
show that the velocity fluctuations are suppressed
when the system enters a dynamically  ordered state due to a
decrease in the effective shaking temperature \cite{53,56,60,61,62}.
Since the drive required to induce dynamical ordering
increases with increasing pinning density,
specific portions of the ConfG sample become ordered for certain values of the
ac driving amplitude.
This produces an effective temperature gradient across each pinning plaquette, with
the largest effective temperature on the high pinning density side.
In Fig.~\ref{fig:10}, we show snapshots of Voronoi constructions
obtained from the system in Fig.~\ref{fig:9}(a).
Figure ~\ref{fig:10}(a) illustrates the configuration
when the driving amplitude reaches its maximum
magnitude in the $+y$ direction for a sample with $F^{y}_{ac}=0.7$, where
there is a normal negative $x$ direction ratchet effect.
Topological defects are uniformly distributed throughout the sample. 
Figure ~\ref{fig:10}(b) shows the same point in the ac drive cycle for a sample with
$F^{y}_{ac}  = 2.3$, where there is a reversed positive $x$
direction ratchet effect.
The more disordered regions are  correlated with the regions of high pinning density,
where the effective shaking temperature temperature is higher. 
As $F^y_{ac}$ increases further, vortices throughout the sample are able to
dynamically reorder during the portion of the drive cycle at which $|F^{y}_{ac}|$ attains
its maximum value, 
so the effective temperature gradient
becomes spatially flat and the ratchet effect is reduced.
This is shown in Fig~\ref{fig:9}(a) at the highest values of $F^{y}_{ac}$. 

In Fig.~\ref{fig:9}(b) we plot $X_{\rm net}$ versus $F^y_{ac}$
in the ConfG array with $F_p=1.0$ at $B/B_{\phi} = 2.5$.
Here the normal negative $x$ ratchet effect
persists over the larger range $0.5 < F^{y}_{ac} < 2.9$
before the ratcheting switches into the reversed positive $x$ direction
due to the dynamical ordering effects.
The increase in the region over which there is a normal ratchet effect
occurs because when the vortex density is lower,
a larger  external drive must be applied to
induce dynamical ordering in the low pinning density regions.
In Fig.~\ref{fig:9}(c) at $B/B_{\phi} = 2.0$, the ratchet reversals are lost and
there is a normal ratchet effect over the range
$0.5 < F^{y}_{ac}< 3.5$.
There is no longer a reversed ratchet effect at
low $F^{y}_{ac}$ due to the lack of interstitial vortices in the high pinning density
portion of the array.  At magnetic fields that are this low, all of the vortices in the
high pinning density areas are trapped in singly or doubly occupied pinning sites, and
there are no remaining freely flowing vortices that could knock one of the pinned vortices
out of a pinning site and generate fluctuations in the $x$ direction velocity.  As a result,
the high pinning density portion of the sample is effectively frozen and prevents the
vortices in the lower pinning density portions of the sample from translating in the $x$
direction.
It is possible that for ac drives larger than those illustrated in
Fig.~\ref{fig:9}(c), a reversed ratchet effect may appear due to the occurrence of partial
dynamical ordering in the sample.

\begin{figure}
\includegraphics[width=3.5in]{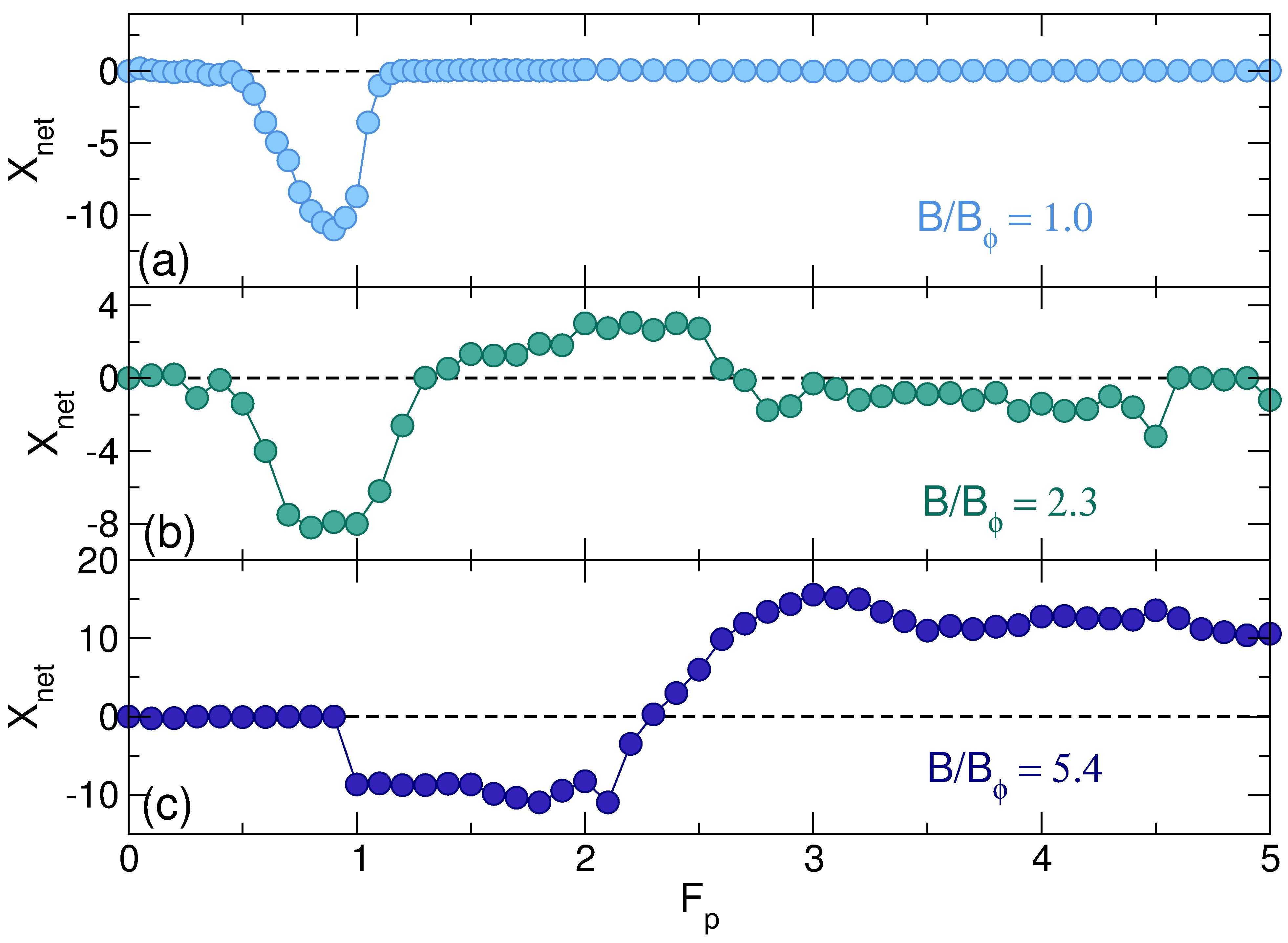}
\caption{ $X_{\rm net}$ after 50 ac drive cycles vs $F_{p}$ for
  ConfG arrays with $F^{y}_{ac} = 0.7$.
  (a) At $B/B_{\phi} = 2.0$, only a normal negative $x$ ratchet effect occurs.
  (b) At $B/B_{\phi} = 2.3$, there are multiple ratchet reversals.
  (c) At $B/B_{\phi}  = 5.4$ there is a reversal from a normal
  negative $x$ to a reverse positive $x$ ratchet effect.
}
\label{fig:11}
\end{figure}

In Fig.~\ref{fig:11}(a) we plot  $X_{\rm net}$ versus $F_{p}$
in a ConfG array with $F^{y}_{ac} = 0.7$ and $B/B_{\phi} = 2.0$.
There is no  
ratchet effect for $F_{p} > 1.2$ since for strong pinning all the vortices 
remain localized at pinning sites during the entire ac drive cycle. 
For $B/B_{\phi} = 2.3$, shown in Fig.~\ref{fig:11}(b),
the ratchet effect is initially in the normal negative $x$ direction for 
$0.5 < F_{p} < 1.2$, and then switches to the reversed positive $x$ direction 
for $1.2 < F_{p} < 2.8$.   Another switch to the normal negative $x$ direction ratchet
occurs at $F_{p}= 2.8$, and the normal ratchet effect gradually diminishes to zero at
the highest values of $F_{p}$.
The onset of the second normal negative $x$ ratchet 
regime is correlated with the appearance
of doubly occupied pinning sites in the high pinning density regions.
These act to reduce the $x$ direction velocity fluctuations of the flowing
interstitial vortices, so that  the effective temperature in the high pinning density
region is smaller than in other portions of the sample.
Figure~\ref{fig:11}(c) shows that at $B/B_{\phi} = 5.4$, there is a large regime of
normal negative $x$ ratchet behavior extending from
$0.9 < F_{p} < 2.3$, followed by a reversed positive $x$ direction ratchet flow
for $F_{p} > 2.3$.  

\subsection{Thermal Fluctuations}

\begin{figure}
\includegraphics[width=3.5in]{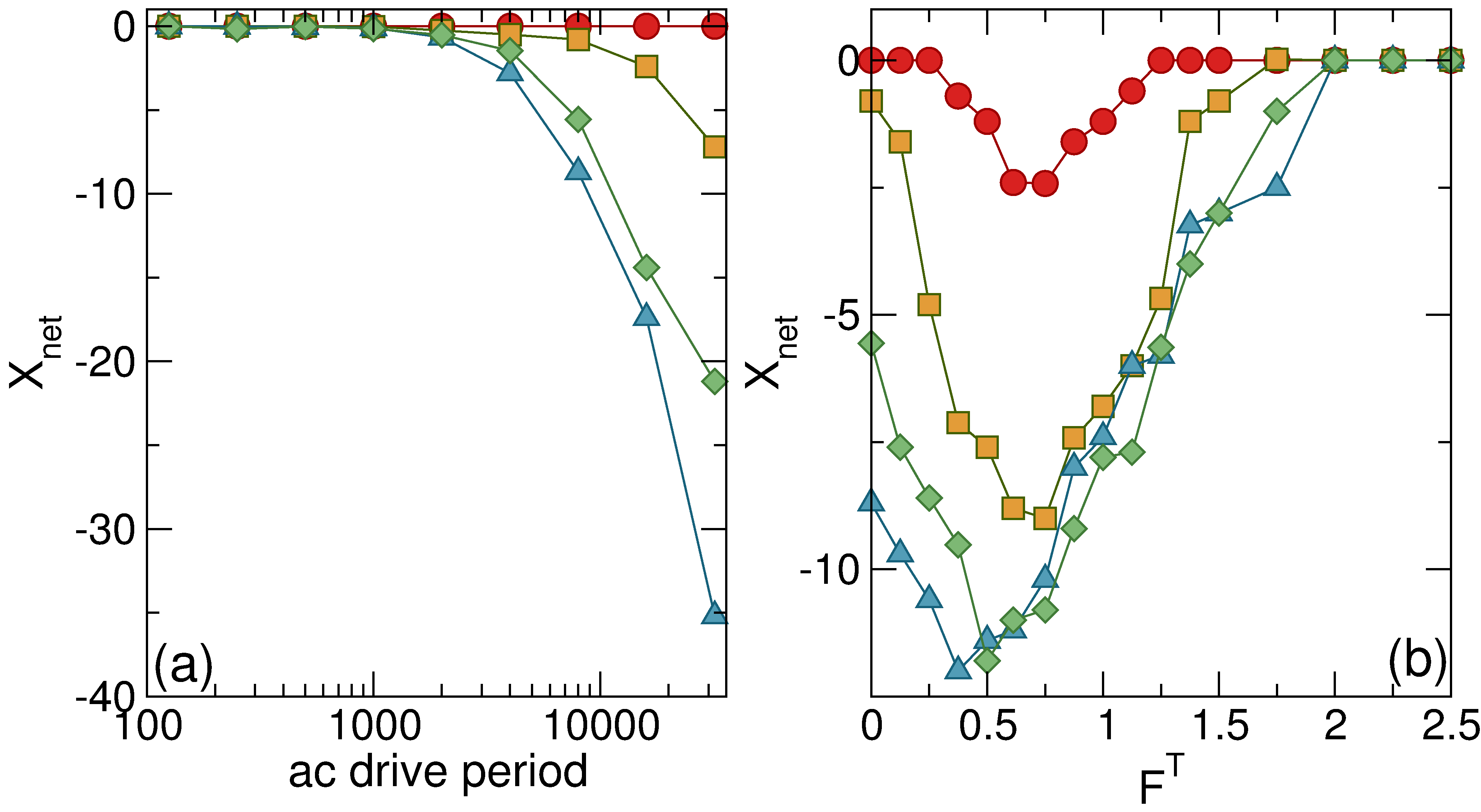}
\caption{
  $X_{\rm net}$ after 50 ac drive cycles
  in a ConfG array with 
  $F_{p} = 1.0$ and $F^{y}_{ac} = 0.7$ at
  $B/B_{\phi} = 0.25$ (red circles), $0.5$ (orange squares), $1.0$ (green diamonds)
  and $1.5$ (blue triangles).
  (a) $X_{\rm net}$ versus ac drive period
  in simulation time steps for $F^{T}=0$, where $F^{T}$ is the amplitude
  of the thermal force term.
  (b) $X_{\rm net}$ vs $F^{T}$
  at an ac drive period of 8000 simulation time steps.
}
\label{fig:12}
\end{figure}

Although thermal fluctuations alone do not produce a ratchet effect in the ConfG array
in the absence of an external drive,
they can  enhance the transverse ratchet effect in some cases. 
In Fig.~\ref{fig:12}(b) we plot $X_{\rm net}$
versus $F^{T}$ for a ConfG sample with $F_p=1.0$ and $F^y_{ac}=0.7$,
where $F^{T}$ is the magnitude of the 
fluctuations in the thermal force term 
added to the vortex equation of motion.
At $B/B_{\phi} = 0.25$ and $F^T=0$, all the vortices are 
pinned during the entire ac drive cycle and $X_{\rm net} = 0$;
however, 
as $F^{T}$ increases a finite ratchet effect emerges
that exhibits a maximum
efficiency at $F^{T} = 0.75$ 
before dropping back to zero for higher values of $F^{T}$.
At $B/B_{\phi} = 0.5$, there is a weak
ratchet effect at $F^{T} = 0$ which undergoes more than a tenfold increase in
magnitude for
increasing $F^{T}$, reaching its
maximum efficiency near $F^{T} = 0.75$.  
At $B/B_{\phi} = 1.0$ there is a robust ratchet
effect at $F^{T} = 0$ which shows 
a small enhancement in magnitude to a maximum efficiency
at $F^{T} = 0.5$ before dropping to  zero at $F^{T} = 2.0$.
For $B/B_{\phi} = 1.5$ there is a similar trend,
with a maximum efficiency at   $F^{T} = 0.6$. 
For higher values of $F^{y}_{ac}$ where the vortices spend more time in motion during
each drive cycle,
the addition of  thermal effects generally decreases the ratchet efficiency.
These results show that when thermal effects are important, the
transverse ratchet effect remains robust and can even be enhanced.
 
In Fig.~\ref{fig:12}(a) we examine the effects of
changing the ac drive period for the same
system in Fig.~\ref{fig:12}(b) at $F^{T} = 0$.
All of the results presented so far were obtained with an ac drive period of
$8000$ simulation time steps.
For $B/B_{\phi} = 0.25$, the vortices are pinned
during the entire ac drive cycle so that $X_{\rm net} = 0$
independent of the value of the ac period. 
For $B/B_{\phi} = 0.5$, 1.0, and $1.5$,  
the ratchet is weak at small ac drive periods since
each vortex simply moves back and forth
within its local potential minimum, so that there is no generation
of the plastic flow necessary 
for the transverse ratchet effect to occur.
As the ac drive period increases, the ratchet effectiveness increases linearly.
This indicates that low frequency ac drive cycles
produce stronger transverse ratchet effects. 

\subsection{Comparison To Longitudinal Gradient Ratchets}

\begin{figure}
\includegraphics[width=3.5in]{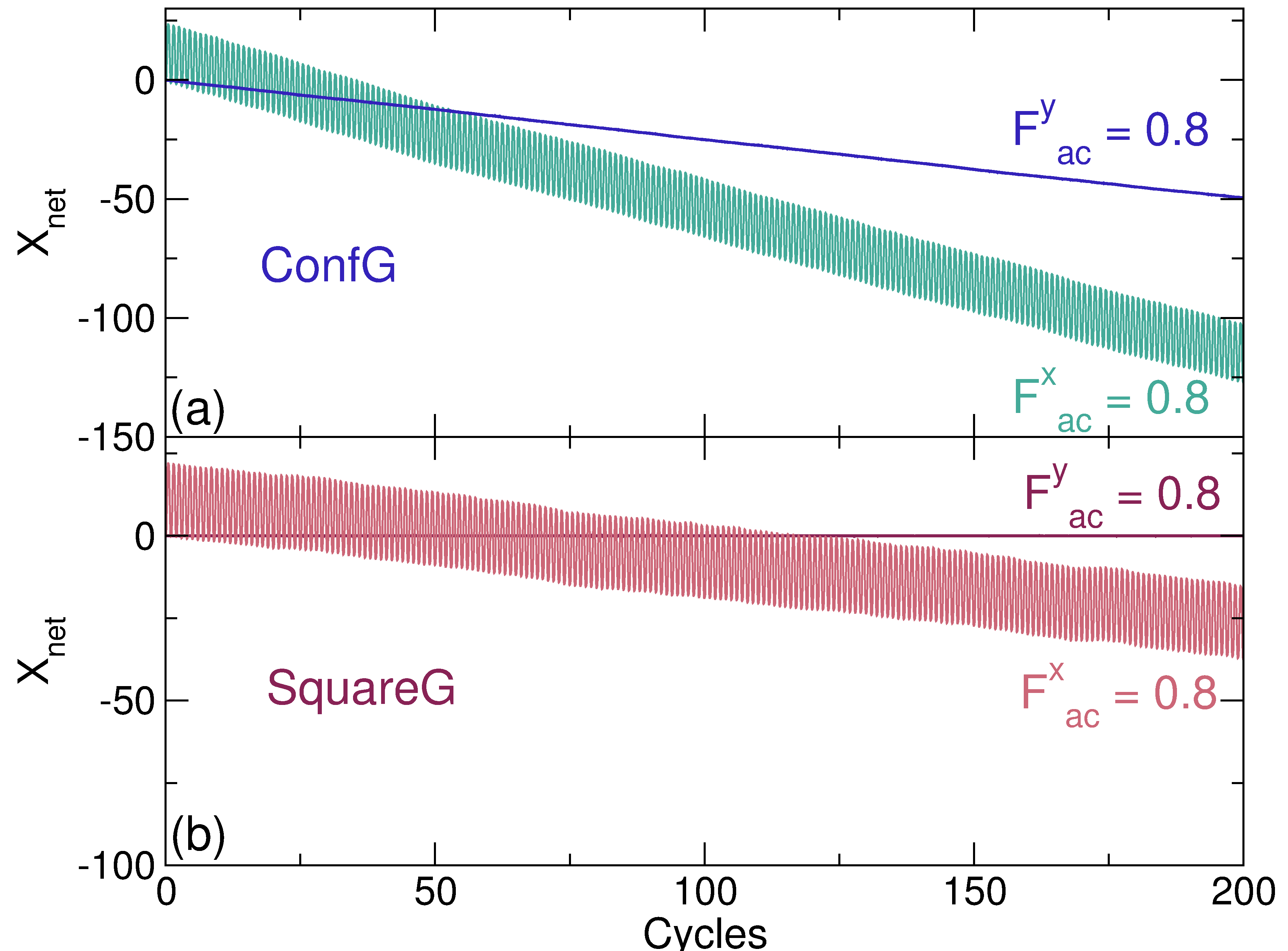}
\caption{$X_{\rm net}$ vs time in ac drive cycles for samples with
$F_{p} = 1.0$ and $B/B_{\phi} = 1.0$.
  (a) ConfG array for $y$-direction driving of $F^y_{ac}=0.8$ (upper
  right blue curve, transverse
  ratchet), and $x$-direction driving of $F^x_{ac}=0.8$
  (lower right green curve, longitudinal ratchet).
  (b)  SquareG array for $y$-direction driving of $F^y_{ac}=0.8$
  (upper right red curve, transverse
  ratchet effect is absent), and $x$-direction driving of $F^x_{ac}=0.8$
  (lower right pink curve,
  longitudinal ratchet).
}
\label{fig:13}
\end{figure}

In previous work, we applied an ac drive in the $x$-direction, parallel to
the substrate asymmetry direction, and demonstrated the existence of
a longitudinal ratchet effect in the  ConfG
array by measuring $X_{\rm net}$ \cite{52}.
In Fig.~\ref{fig:13}(a)  we show the time evolution of
$X_{\rm net}$ during $200$ ac drive cycles for a ConfG sample with
$B/B_{\phi} = 1.0$ and $F_{p} = 1.0$ for
$F^{y}_{ac} = 0.8$, where a transverse ratchet effect appears, and
for
$F^{x}_{ac} = 0.8$, where there is a longitudinal ratchet effect.
There are strong oscillations in $X_{\rm net}$ for
the longitudinal ratchet that arise because the ac drive direction is the same as
the ratchet motion direction.
The longitudinal ratchet effect is approximately 2.5
times larger than the transverse ratchet effect,
and we find that in general the longitudinal
ratchet is more effective than the transverse ratchet
by a similar ratio for other parameters. 
The longitudinal 
ratcheting is stronger since it is a rocking ratchet effect. 
In contrast, in the transverse ratchet effect the ac drive does not directly
push the vortices in the direction of the asymmetry  but instead
generates plastic flow, which creates the transverse
velocity fluctuations that permit a
correlation ratchet effect to occur.
We find a similar ratio of the
longitudinal to transverse ratchet effects for the RandG arrays as well (not shown);
however, the overall ratchet effect is smaller in each case for the RandG array than for
the ConfG array.
In Fig.~\ref{fig:13}(b) we plot $X_{\rm net}$ versus time
for a SquareG array under the same $x$ and $y$ ac driving conditions.
Here there is no transverse ratchet effect, but there is still a longitudinal ratchet effect
which is about four times
less effective than the longitudinal ratchet effect for the ConfG array.
In general we find that the transverse ratchet effect
is more sensitive to changes in magnetic field than the longitudinal ratchet effect
since commensuration effects
strongly influence the magnitude of the dynamical fluctuations responsible
for the transverse ratchet effect.

\section{Drift Ratchet}

\begin{figure}
\includegraphics[width=3.5in]{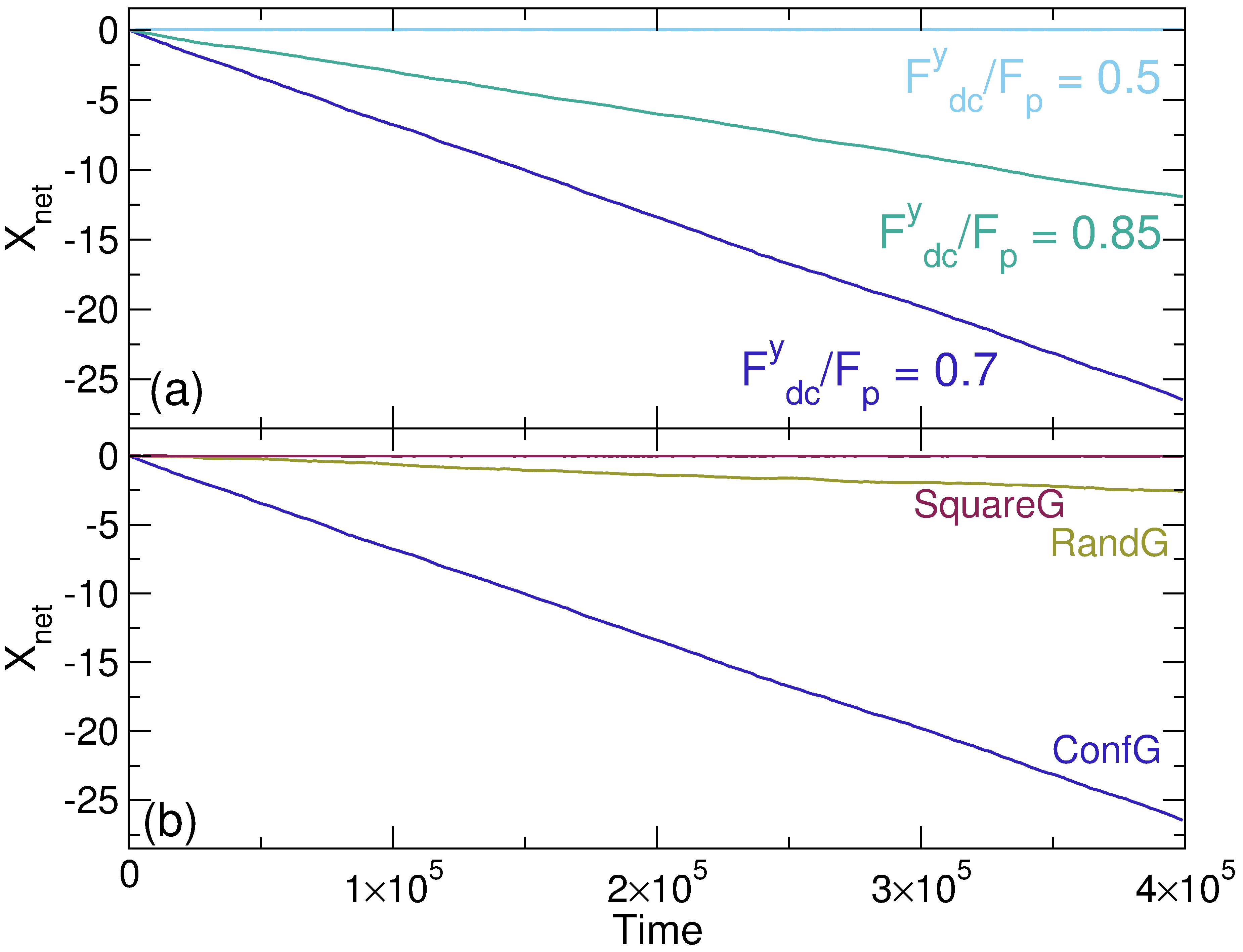}
\caption{ $X_{\rm net}$ vs time in simulation time steps
  for arrays with a dc drive applied in the $y$ direction to create a
  geometric ratchet effect in samples with $B/B_{\phi} = 1.0$ and
  $F_{p} = 1.0$.
  (a) ConfG array at $F^{y}_{dc} = 0.5$ (upper light blue curve) where there is no drift,
  at $F^{y}_{dc} = 0.7$ (lower dark blue curve) where there is strong drift,
  and at $F^{y}_{dc} = 0.85$ (center green curve) where there is a reduced drift.
  (b) The ConfG (lower dark blue curve), RandG (center brown curve), and SquareG
  (upper red curve) arrays at  $F^{y}_{dc} = 0.7$.  The SquareG array shows no drift while
  the RandG array has a
  transverse drift that is approximately 10 times smaller than that
  of the ConfG array.  
}
\label{fig:14}
\end{figure}

We next consider the case where instead of an ac drive, we apply
a dc drive in the $y$-direction, and we measure the net drift of vortices in the
$x$-direction to examine the drift or geometric ratchet effect. 
In Fig.~\ref{fig:14}(a) we plot $X_{\rm net}$ versus time
in simulation time steps for the ConfG array
at $F_{p} = 1.0$, $B/B_{\phi} = 1.0$, and varied $F^y_{dc}$. 
For $F^{y}_{dc} = 0.5$, although there is flow in the $y$-direction,
there is no drift of the vortices in the $x$-direction, 
while at $F^{y}_{dc} = 0.7$ there is
a pronounced $x$-direction drift,
with individual vortices moving
an average of $25\lambda$
in the negative $x$ direction after $4 \times 10^5$ simulation time steps.
In comparison,
for an ac drive of  $F^{y}_{ac} = 0.7$ during the same amount of time
(equivalent to 50 ac drive cycles), Fig.~\ref{fig:3}(a) indicates that
individual vortices move an average distance of only $8\lambda$
in the negative $x$ direction, showing
that the transverse drift ratchet is 
approximately three times more effective at transporting the vortices
than the ac transverse ratchet effect.
For $F^{y}_{dc} = 0.85$, Fig.~\ref{fig:14}(a) shows that the 
transverse drift is reduced. 
In Fig.~\ref{fig:14}(b) we plot $X_{\rm net}$ versus time at
$F^{y}_{dc} = 0.7$ for the SquareG, RandG, and ConfG arrays.
There is no transverse drift in the SquareG array
since the vortices move in predominately straight trajectories along the
$y$-direction.   The RandG array shows a transverse drift 
that is approximately 10 times smaller than the transverse drift
in the ConfG array.

\begin{figure}
\includegraphics[width=3.5in]{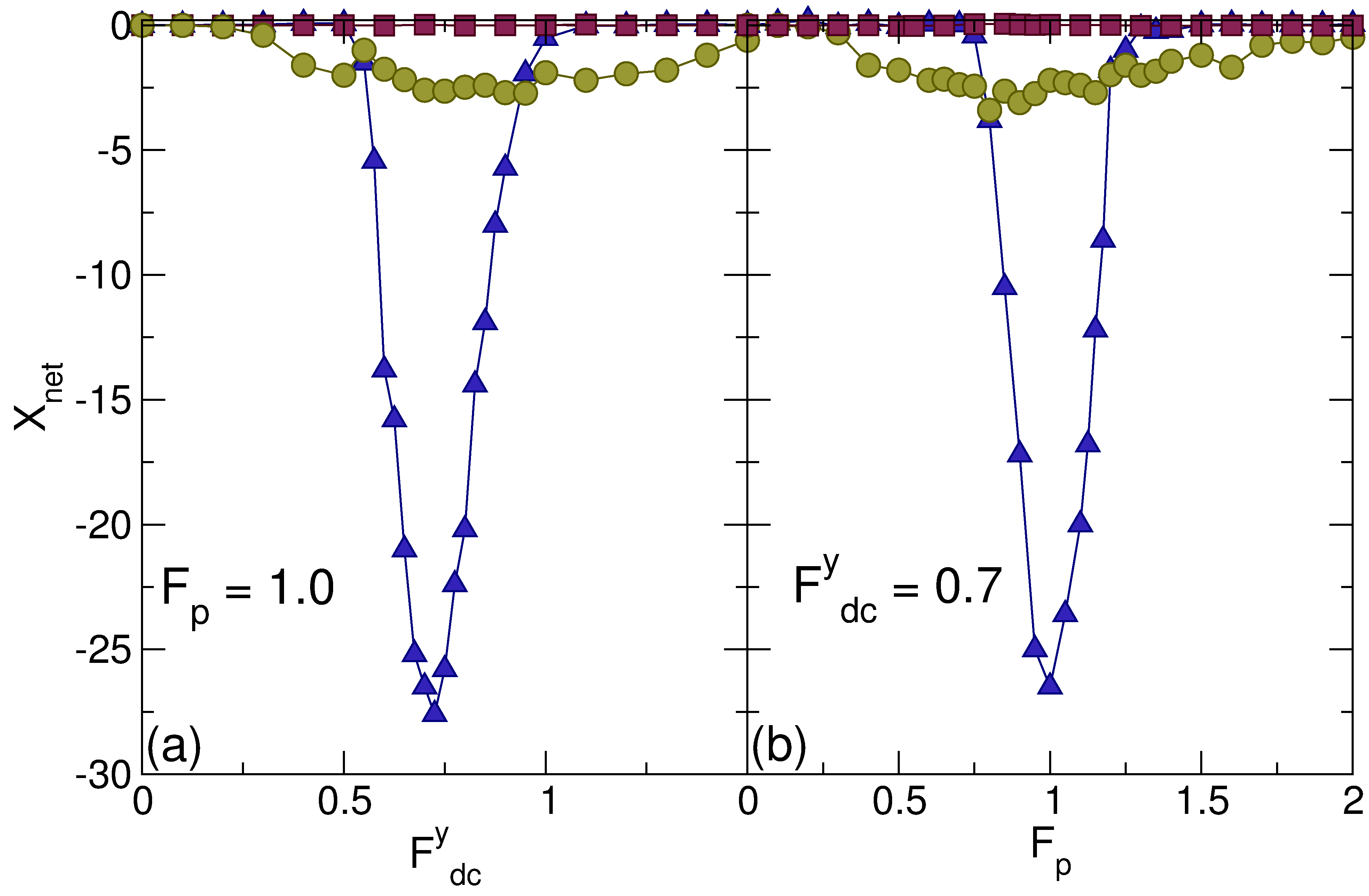}
\caption{
  (a) $X_{\rm net}$ after $4 \times 10^5$ simulation time steps
  vs $F^{y}_{dc}$ for
  ConfG (blue triangles), RandG (brown circles), and SquareG (red squares)
  arrays with $B/B_{\phi} = 1.0$ and $F_{p} = 1.0$.
  (b) $X_{\rm net}$ vs $F_{p}$ for
  ConfG (blue triangles), RandG (brown circles), and SquareG (red squares) arrays
  with $B/B_{\phi} = 1.0$ and $F^y_{dc} = 0.7$.
}
\label{fig:15}
\end{figure}

In order to compare to the ac driven results, we measure
$X_{\rm net}$ for the dc driven system
after $4\times10^5$ simulation time steps, which corresponds to 
the same time interval required to complete 50 ac drive cycles in the
ac driven system with a period of 8000 simulation time steps.
In Fig.~\ref{fig:15}(a) we plot $X_{\rm net}$ versus
$F^{y}_{dc}$ for ConfG, RandG, and SquareG arrays with
$B/B_{\phi} = 1.0$ and $F_{p} = 1.0$.
The transverse drift is strongest for the ConfG array
and rapidly drops off when $F^{y}_{dc} > F_{p}$.
In the ac driven systems, strong ratchet effects can persist
for $F^{y}_{ac} > F_{p}$ since there
is a portion of the ac cycle during which
the driving force  is smaller than $F_{p}$ so that plastic flow can occur. 
In the dc driven case, however, when $F^{y}_{dc} > F_{p}$
all the vortices are moving at all times and the plastic flow necessary
to produce the correlation ratchet effect is lost.
The transverse drift ratchet is smaller in the RandG array
but persists over a wider range of values of  $F^{y}_{dc}$  due to the stronger
dispersion in the pinning forces for the random array caused by overlap of pinning
sites in some locations, which creates
local regions where  the effective pinning force is larger than $F_{p} = 1.0$.  
In the SquareG array, the vortex trajectories are one-dimensional along the $y$ direction
for all values
of $F^{y}_{dc}$, so there is no transverse drift ratchet effect.
Figure~\ref{fig:15}(b) shows $X_{\rm net}$ for the same samples plotted against
$F_p$ at fixed $F^{y}_{dc} = 0.7$.
The transverse drift in the ConfG and RandG arrays
is lost for low pinning forces when the vortex flow is elastic, as well as at
large pinning forces where all the vortices remain pinned.
If we apply the external dc drive in the negative $y$-direction (not shown),
we find exactly the same drift ratchet effects, with the vortices still moving
in the negative $x$-direction. 

The drift ratchet effect we observe is different in nature from
transverse drift ratchets studied by other groups \cite{7,8,9,10,11}. 
In those systems, the
particle-particle interactions are not important and the ratchet effect arises when
particles are deflected during collisions
with obstacles or pinning sites, producing a net drift.
In contrast, the transverse drift ratchet effect we observe is produced by transverse
nonequilibrium fluctuations generated by particle-particle interactions, leading to
the emergence of an effective noise correlation ratchet.

\subsection{Drift Ratchet Reversals}

\begin{figure}
\includegraphics[width=3.5in]{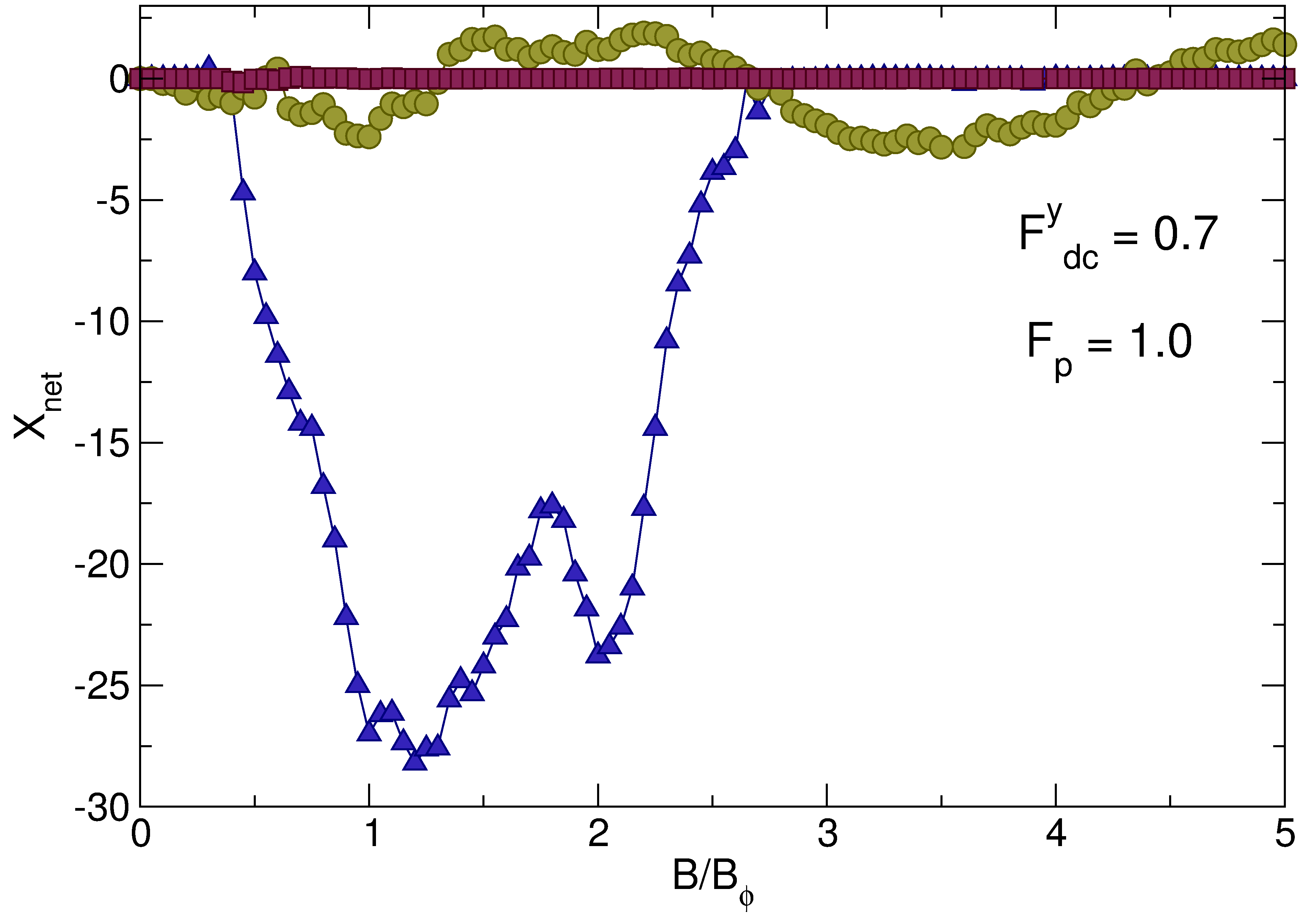}
\caption{$X_{\rm net}$ after $4 \times 10^5$ simulation time steps vs $B/B_{\phi}$ for
  ConfG (blue triangles), RandG (brown circles), and SquareG (red squares) arrays
  at $F_{p} = 1.0$ and $F^{y}_{dc} = 0.7$.
The RandG array shows multiple reversals in the
direction of the transverse drift, the SquareG array shows no transverse drift, and the
ConfG array has the largest transverse drift magnitude.  
}
\label{fig:16}
\end{figure}

In previously studied drift ratchets, reversals in the direction of drift were not observed
\cite{7,8,9,10,11,56}.
In the drift ratchet described here,
there can be reversals of the drift direction in both the ConfG and RandG arrays.
In Fig.~\ref{fig:16} we plot 
$X_{\rm net}$ versus $B/B_{\phi}$ for ConfG, RandG, and SquareG arrays
with $F_{p} = 1.0$ and $F^{y}_{dc} = 0.7$.
The conformal array shows a negative drift ratchet effect with local efficiency maxima
at $B/B_{\phi} = 1.25$ and
$B/B_{\phi}=2.0$,  while the drift is suppressed for $B/B_{\phi} > 3.0$.
In comparison, in the RandG array the magnitude of the drift is smaller 
but there are multiple reversals from a normal negative $x$ to a
reversed positive $x$ drift ratchet effect.
The SquareG array shows no transverse drift.
 
\begin{figure}
\includegraphics[width=3.5in]{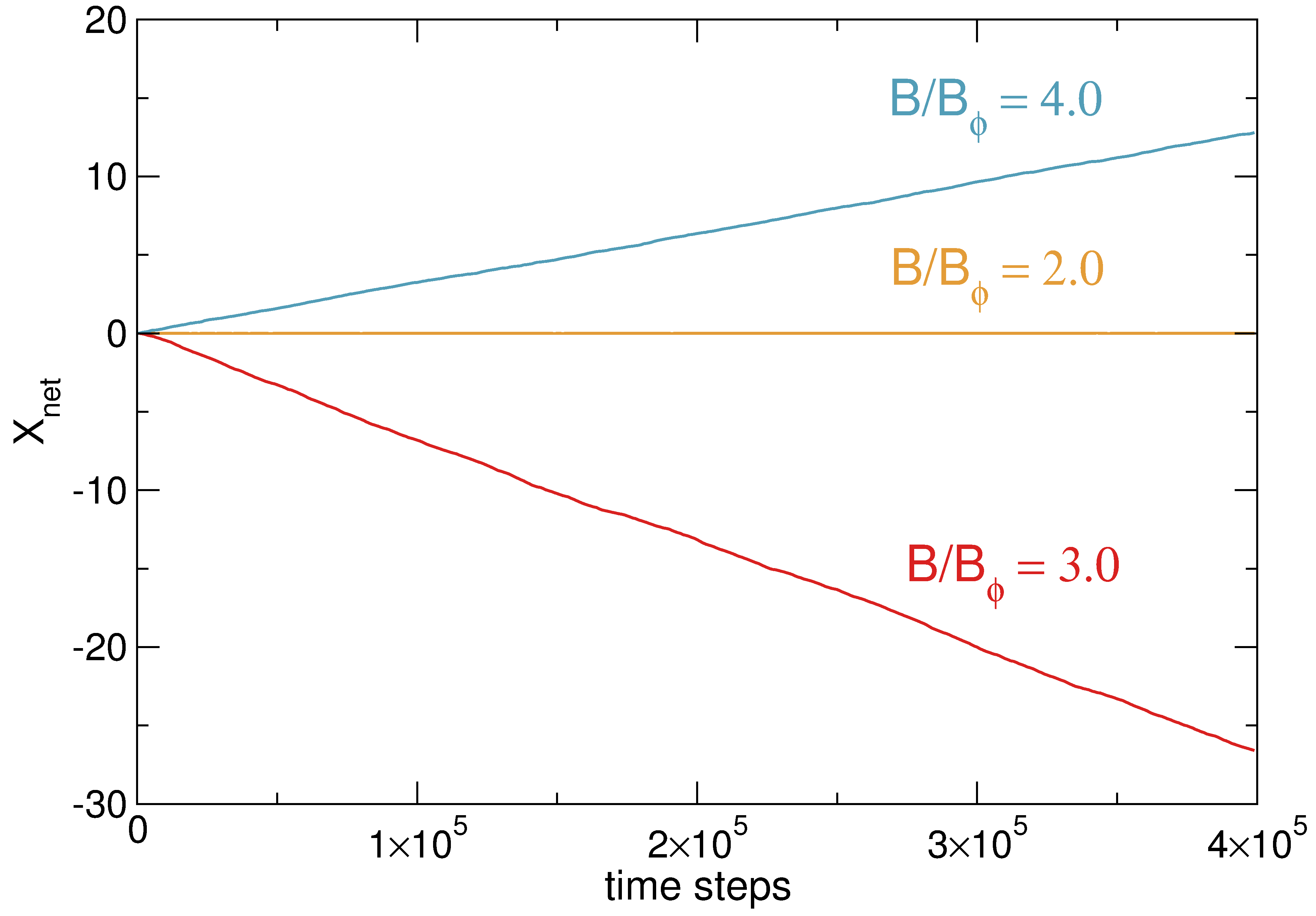}
\caption{ $X_{\rm net}$ vs time in simulation time steps for a ConfG array with
  $F_{p} = 3.0$ and $F^{y}_{dc} = 0.9$.
  Middle yellow curve: $B/B_{\phi} = 2.0$, where there is no transverse drift.
  Lower red curve: $B/B_{\phi} = 3.0$, where there is a normal negative $x$ transverse
  drift.
  Upper blue curve: $B/B_{\phi} = 4.0$, where there is a
  reversed positive $x$ transverse drift,
  indicating that there is a reversal in the conformal drift ratchet direction as a
function of magnetic field.  
}
\label{fig:17}
\end{figure}

For the ConfG array,  drift ratchet reversals  generally occur for stronger pinning
and fillings of $B/B_{\phi} > 2.0$. 
In Fig.~\ref{fig:17} we plot $X_{\rm net}$ versus time in a ConfG array
with $F_{p} = 3.0$ and $F^{y}_{dc} = 0.9$.
At $B/B_{\phi} = 2.0$ 
there is no transverse drift effect,
while at $B/B_{\phi} = 3.0$ there is a normal negative $x$ drift and
at $B/B_{\phi} 4.0$ there is a reversed positive $y$
drift, indicating a reversal in the drift direction
as a function of magnetic field.
We again find that the magnitude of the transverse drift for the dc driven systems
is significantly larger than the ac driven transverse ratchet effect. 

\begin{figure}
\includegraphics[width=3.5in]{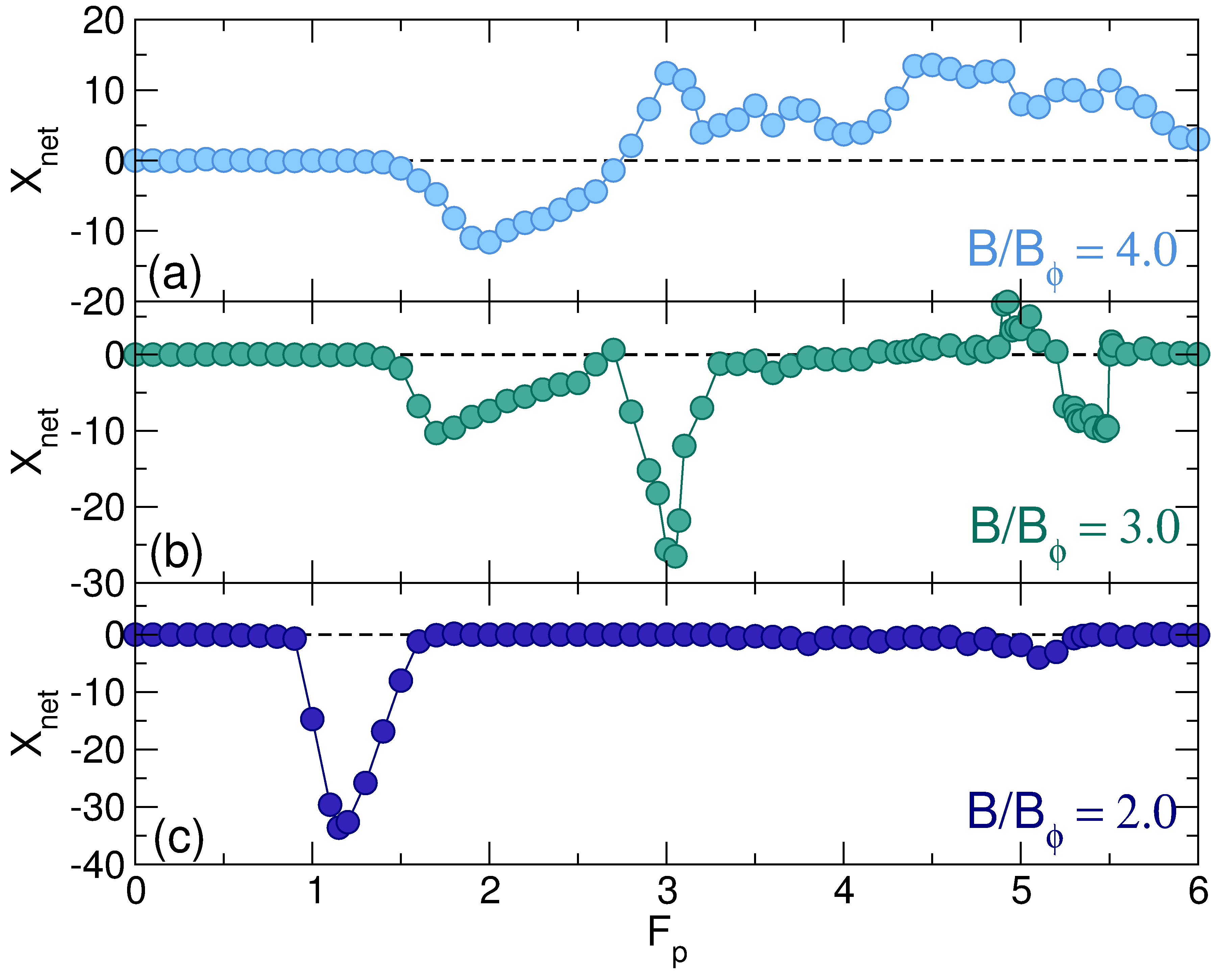}
\caption{ $X_{\rm net}$ after $4 \times 10^5$ simulation time steps vs $F_{p}$
  for the ConfG system in Fig.~\ref{fig:17} with $F^{y}_{dc} = 0.9$.
  (a) At $B/B_{\phi} = 4.0$ there is a reversal from normal negative $x$ to
  reversed positive $x$ drift.
  (b) At $B/B_{\phi} = 3.0$ the drift is mostly in the normal negative $x$ direction.
  (c) At $B/B_{\phi} = 2.0$ the drift is always in the normal negative $x$
  direction and has the largest
  magnitude in the range $0.8 < F_{p} < 1.6$.  
}
\label{fig:18}
\end{figure}

In Fig.~\ref{fig:18}(a) we plot $X_{\rm net}$ after $4 \times 10^5$ simulation time
steps versus $F_{p}$ for the ConfG array from Fig.~\ref{fig:17} with
$F^{y}_{dc} = 0.9$ and $B/B_{\phi} =  4.0$.
For $F_p<1.5$, $X_{\rm net}=0$, while the drift is in the normal negative $x$ direction for
$1.5 < F_{p} < 2.8$ and in the reversed positive $x$ direction for $F_{p} > 2.8$.
There are several local extrema in the drift which correspond to changes in the vortex flow. 
For $B/B_{\phi} = 3.0$, shown in Fig.~\ref{fig:18}(b),
the drift is mostly in the normal negative $x$ direction
with only a small region of  reversed positive $x$ direction drift near
  $F_{p} = 5.0$.
There are also several local extrema in the drift near $F_{p} = 1.6$, 2.7,
and $3.0$.
For $B/B_{\phi} = 2.0$  in Fig.~\ref{fig:18}(c), the drift is always
in the normal negative $x$ direction  and is
largest over the range
$0.8 < F_{p} < 1.6$.
There are some  small fluctuations in the drift near $F_{p} = 5.0$
which corresponds to the point at which the pinning is strong enough that
almost all of the pinning
sites are doubly occupied.

\begin{figure}
\includegraphics[width=3.5in]{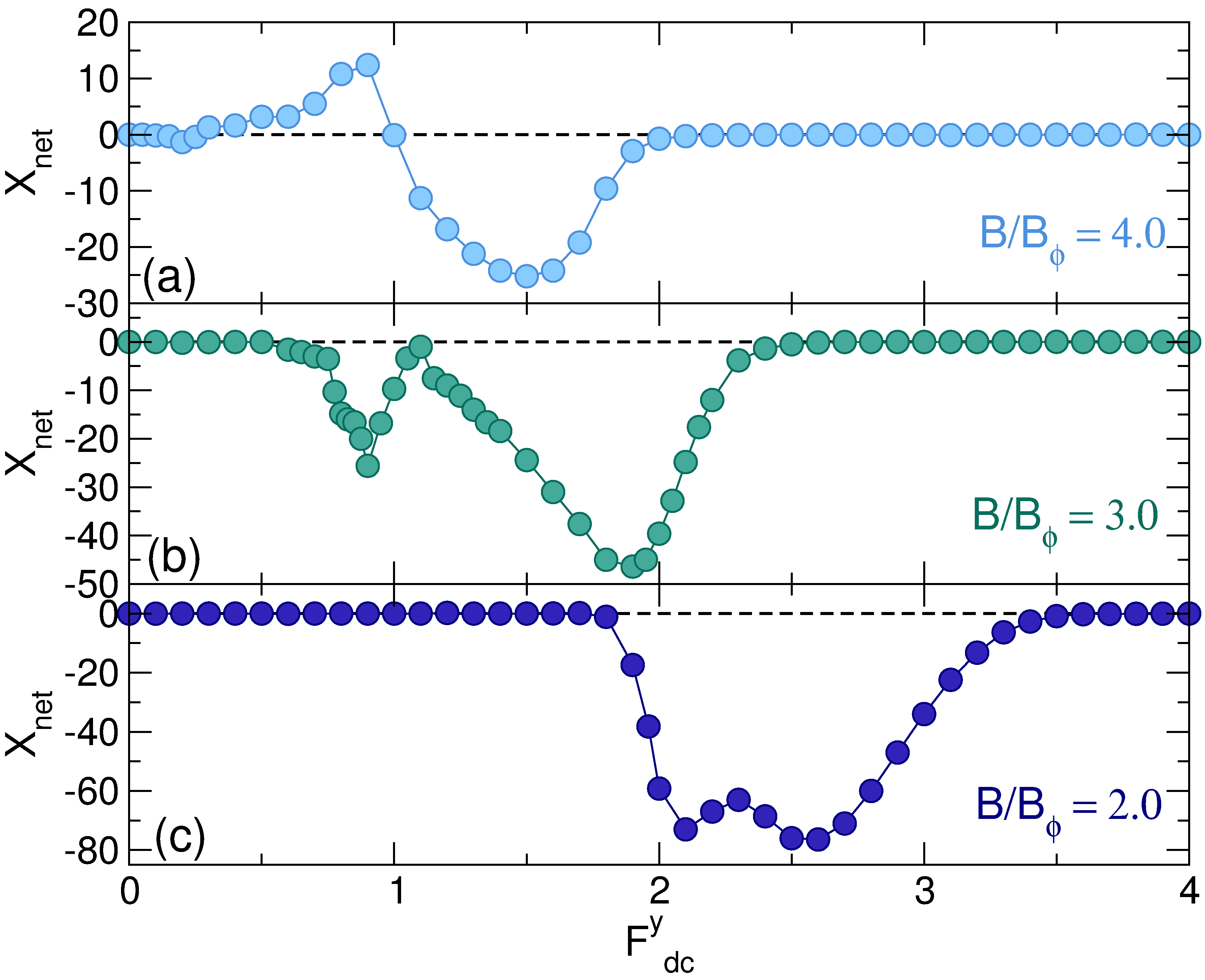}
\caption{ 
  $X_{\rm net}$ after $4 \times 10^{5}$ simulation time steps
  vs $F^{y}_{dc}$ for the ConfG array at $F_{p} = 3.0$.
  (a) At $B/B_{\phi} = 4.0$ there is a reversal in the drift direction.
 (b) At $B/B_{\phi} = 3.0$ the drift is in the normal negative $x$ direction. (c)
  At $B/B_{\phi} = 2.0$, a strong
  normal negative $x$ direction drift occurs for $ 1.9<F^{y}_{dc} < 3.4$.  
}
\label{fig:19}
\end{figure}

Ratchet reversals can also occur as a function of the dc drive magnitude.
In Fig.~\ref{fig:19}(a) we plot $X_{\rm net}$ 
versus $F^{y}_{dc}$ for a ConfG array
at $F_{p} = 3.0$ and $B/B_{\phi} = 4.0$.
There is a reversed positive $x$ direction drift for $ 0.25 < F^{y}_{dc} < 1.0$ 
which is correlated with 
all the pinning sites in the sample being doubly occupied.
The interstitial vortices in the high pinning density regions are
close enough to the doubly occupied pins to cause vortices
to depin, while in the low pinning density regions
the interstitial vortices move around the occupied pinning sites and do not
induce any vortex depinning.  As a result,
the transverse fluctuations are largest in  the
high pinning density portions of the sample, and
the drift motion is in the reversed positive $x$ direction.
For $F^{y}_{dc} > 1.0$, the drive is large enough that all 
the vortices in  doubly occupied pinning sites can be depinned,
producing drift in the normal negative $x$ direction, while
for $F^{y}_{dc} > 2.0$ the transverse velocity
fluctuations become homogeneous throughout the sample and the ratchet 
effect is lost.
Figure~\ref{fig:19}(b) shows that at $B/B_{\phi} = 3.0$,  there are no longer enough
interstitial vortices
in the high pinning density regions of the sample to easily depin vortices
from the  doubly occupied sites, so only a normal negative $x$ direction drift appears.
In Fig.~\ref{fig:19}(c), at $B/B_{\phi} = 2.0$ there is a normal negative
$x$ direction drift ratchet
effect only in the range
$1.9<F^{y}_{dc} < 3.4$

\subsection{Transverse Velocity Fluctuations}

\begin{figure}
\includegraphics[width=3.5in]{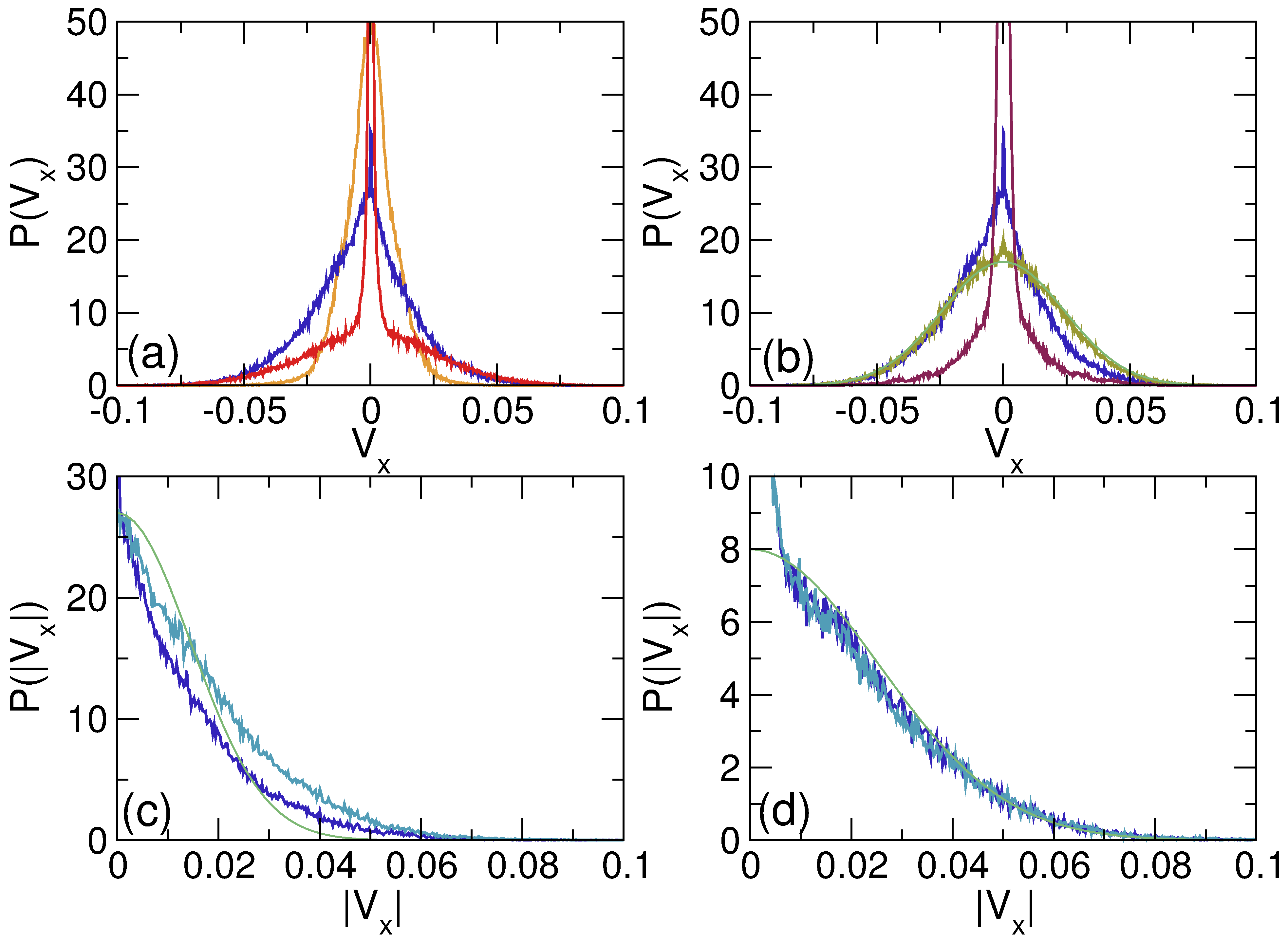}
\caption{ 
  (a) $P(V_x)$ for the ConfG drift ratchet at $B/B_{\phi} = 1.0$ and $F_{p} = 1.0$ for
  $F^{y}_{dc} = 0.55$ (red curve),
  $F^{y}_{dc} = 0.7$ (blue curve) where the strongest transverse drift occurs, and
  $F^{y}_{dc} = 0.95$ (yellow curve).   
  (b) $P(V_x)$ for the ConfG (blue curve),
  RandG (brown curve), and SquareG (dark red curve) drift ratchets
  at $B/B_{\phi}= 1.0$, $F_{p} = 1.0$,
  and $F^{y}_{dc} = 0.7$.
 The smooth green line indicates a Gaussian fit.
  (c)
  $P(|V_{x}|)$ for the ConfG drift ratchet at $B/B_{\phi}=1.0$, $F_p=1.0$, and
  $F^{y}_{dc} = 0.7$.
  The upper light blue curve is for negative $V_x$ values, the
  lower dark blue curve is for positive $V_x$ values, and
the smooth green line is a Gaussian fit.
(d)
  $P(|V_{x}|)$ for the ConfG drift ratchet at $B/B_{\phi}=1.0$, $F_p=1.0$, and
  $F^{y}_{dc} = 0.55$, where there is no net drift.
  The light blue curve is for negative $V_x$ values, the
  dark blue curve is for positive $V_x$ values, and
the smooth green line is a Gaussian fit.
There is little to no asymmetry in the velocity curves, which are both well fit
by a Gaussian distribution away from the peak near $|V_x|=0$.   
}
\label{fig:20}
\end{figure}

We next consider the drift ratchet 
transverse velocity fluctuations measured in the same way as for the
ac driven transverse ratchet effect in Section III.
In Fig.~\ref{fig:20}(a) we plot $P(V_{x})$ for the
ConfG array at $B/B_{\phi} = 1.0$ and $F_{p} = 1.0$.
At $F^{y}_{dc} = 0.55$, where there is no 
transverse drift, 
there is a strong peak in $P(V_x)$
at $V_{x} = 0$ since a portion of the vortices are permanently  pinned.  
For $F^{y}_{dc} = 0.7$ where there is a strong
transverse drift ratchet effect,
the magnitude of the $V_x=0$ peak in
$P(V_{x})$ is reduced since the vortices are only temporarily rather than
permanently pinned, while
$P(V_x)$ remains large over a wider range of $V_x$ values.
At $F^{y}_{dc} = 0.95$ 
where the drift effect is diminished, the width of $P(V_{x})$  
is strongly reduced since the vortices are moving primarily along
straighter trajectories aligned with the $y$-direction.
In Fig.~\ref{fig:20}(b) we plot $P(V_x)$ for the ConfG, RandG, and SquareG arrays
at $B/B_{\phi}=1.0$, $F_p=1.0$, and
$F^{y}_{dc} = 0.7$.
The SquareG array has
a strong peak in $P(V_x)$ at $V_{x} = 0$ since the
vortices are moving in 1D paths in the $y$-direction. 
For the RandG array the velocity fluctuations can be fit to a Gaussian curve as
indicated by the smooth solid line.   
When there is a strong transverse  drift
ratchet effect, $P(V_{x})$ is  asymmetric  about $V_x=0$, as shown in
Fig.~\ref{fig:20}(c) where we plot $P(|V_{x}|)$ separately for 
$V_x>0$ and $V_x<0$ in the ConfG array at $B/B_{\phi}=1.0$, $F_p=1.0$, and
$F^{y}_{dc} = 0.7$. 
There is a clear difference in the velocity distributions for vortices moving in the positive
and negative $x$ directions.
The smooth curve is a Gaussian fit highlighting the non-Gaussian nature of the fluctuations. 
In Fig.~\ref{fig:20}(d) we show $P(|V_x|)$ for positive and negative $V_x$ in the
same ConfG array at  $F^{y}_{dc} = 0.55$ where there is no net drift.  Here the
velocity distribution is symmetric and can be fit to a Gaussian tail away from $|V_x|=0$.

\section{Ratchet Effects for Colloidal Particles}

The results we find should be general to a wide class of systems of interacting 
particles moving over a gradient substrate array
where nonequilibrium transverse fluctuations can be dynamically generated.
For example, our results could be applied to charge-stabilized
colloids interacting with optical trap arrays \cite{57,63}.
Charged colloids can be modeled as overdamped particles
interacting via a repulsive Yukawa or screened Coulomb potential 
$V(R_{ij}) = A_{c}\exp(-\kappa R_{ij})/R_{ij}$,
where $\kappa$ is the screening length and $A_{c}$ is proportional 
to the effective charge on the particle. 
Compared to superconducting vortices,
the colloids have a much shorter range interaction but a sharper 
repulsion at the shortest distances.
The effective charge and the screening length can be tuned readily in the colloidal
system by modifying the ion concentration in the solution.
If transverse ratchets can be realized in a colloidal system, they could potentially
provide a new spatial separation technique in which
a mixture of colloidal species driven over a substrate has
one species 
gradually move further in the drift direction than the other.
In previous studies of the longitudinal ratchet effect on
conformal substrates,
colloidal particles exhibited a robust ratchet effect similar to the
superconducting vortices \cite{52}. 

\begin{figure}
\includegraphics[width=3.5in]{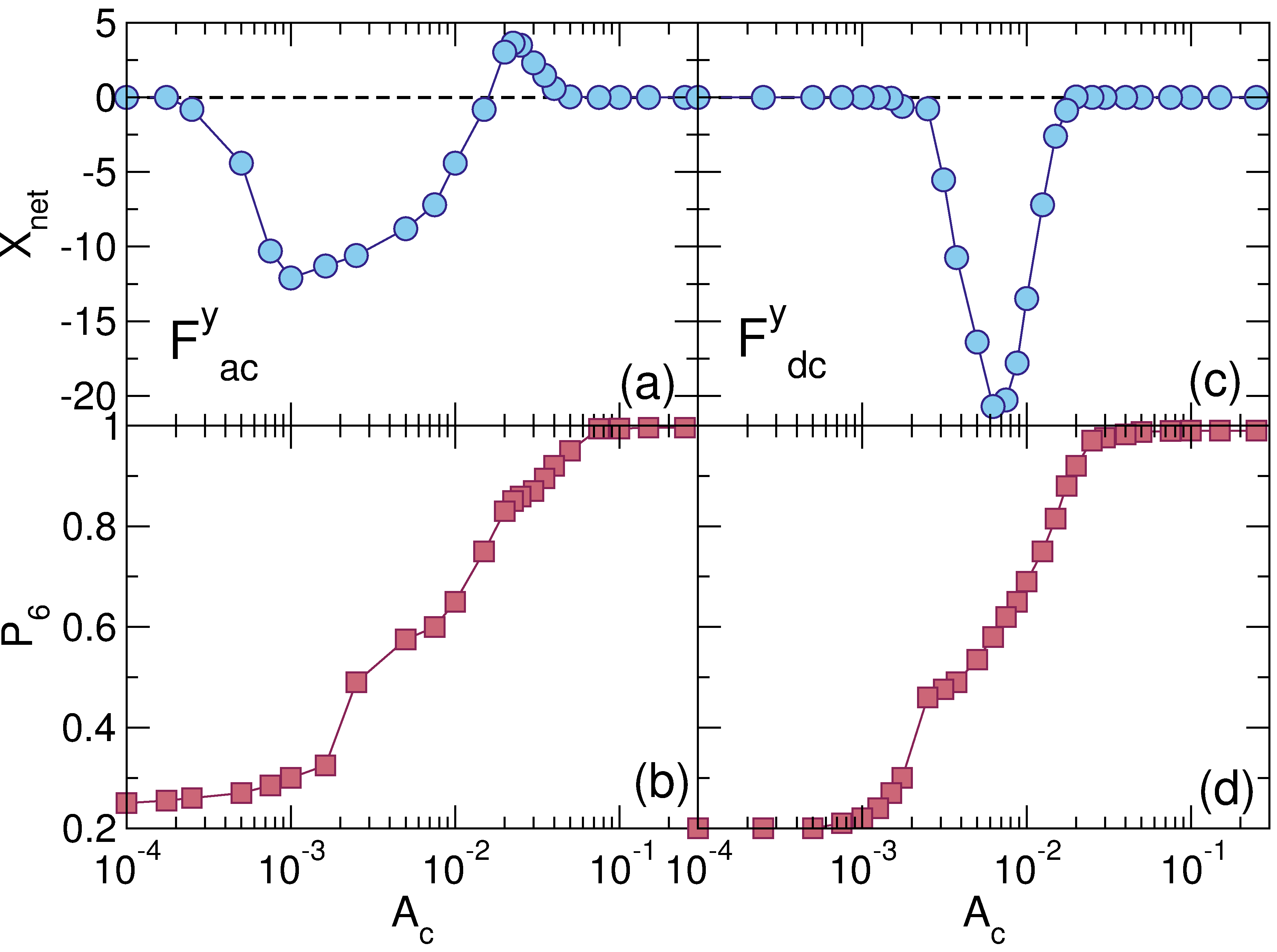}
\caption{ (a) $X_{\rm net}$ after 50 ac drive cycles vs $A_{c}$ for colloidal particles
  interacting with a ConfG array with $F_{p} = 1.0$, a filling factor of $1.0$,
  and $F^{y}_{ac} = 0.7$.
  Here $A_{c}$ is the prefactor  of the colloidal repulsive interaction term.
  (b) The corresponding fraction of six-fold coordinated colloids $P_6$ vs $A_{c}$.
  (c) $X_{\rm net}$ after $4 \times 10^5$ simulation time steps
  vs $A_{c}$ for the same system with $F^{y}_{dc} = 0.7$.
(d) The corresponding $P_{6}$ vs $A_{c}$.  
}
\label{fig:21}
\end{figure}

In Fig.~\ref{fig:21}(a) we plot $X_{\rm net}$ versus $A_{c}$
for colloids driven oven a ConfG array with a transverse ac
drive $F^{y}_{ac} = 0.7$ for $F_{p} = 1.0$ at
a filling fraction of $1.0$.  In Fig.~\ref{fig:21}(b) we plot the
corresponding fraction of six-fold coordinated colloids $P_6$ versus $A_{c}$.
For the smallest values of $A_{c}$, the colloids
are weakly interacting and become localized at the pinning sites so that there is
no transverse ratchet.  At intermediate
values of $A_{c}$, where $ 0.3 < P_{6} < 0.82$,
the colloids move in the negative $x$ direction and exhibit a normal
transverse ratchet effect, while at slightly higher values of $A_c$ there is a small
window in which the ratchet 
effect is in the reversed positive $x$ direction and $0.9 < P_{6} < 1.0$.
This reversed ratchet
effect is similar to what we observe for large values of $F^{y}_{ac}$
in the vortex system, and arises when
the colloids dynamically order in the low pinning density portions of the
sample but remain partially
disordered in the high pinning density portions of the sample.
For higher values of $A_{c}$, when the colloids form 
a uniform triangular lattice with
$P_{6} = 1.0$, the flow becomes  elastic and the ratchet effect is lost. 
Figure~\ref{fig:21}(c) shows $X_{\rm net}$
versus $A_c$ for dc driven colloids in the drift ratchet configuration where the drive is
applied along the $y$-direction.
Here there is only a normal negative $x$ direction transverse drift effect, 
and the magnitude of $|X_{\rm net}|$ is larger than that for the ac driven
sample in Fig.~\ref{fig:21}(a).
At high enough 
values of $A_{c}$, the system again forms a triangular solid with
$P_{6} = 1.0$  and the transverse drift is lost. These results show that 
the transverse ratchet and drift ratchet along with
ratchet reversals can also be  realized in colloidal systems.   

\section{Summary}

Using numerical simulations we show that vortices on
conformal pinning arrays driven by an ac force applied perpendicular to
the array asymmetry direction exhibit
a novel transverse ratchet effect where there is a net drift of vortices
perpendicular to the ac drive. 
This effect arises when the
dynamically generated nonequilibrium fluctuations
which have non-Gaussian characteristics combine with the asymmetry of the substrate 
to create what is known as a noise correlation ratchet.
This ratchet effect is 
distinct from previously observed transverse vortex ratchet
effects that are caused by geometric deflection of individual vortex trajectories.
In our system, the correlated noise
is generated by the plastic flow of vortices,
which creates strong non-Gaussian velocity fluctuations
in the direction perpendicular to the ac drive. 
We find that the transverse ratchet effect is absent for square gradient arrays since these
produce less meandering of the vortex trajectories and therefore have reduced velocity
fluctuations.
The effect is present for random gradient arrays but is substantially weaker than that
produced by the conformal array.
We show that it is 
possible to realize a series of reversals in the direction of flow of the
transverse ratchet 
due to changes in the spatial flow pattern of the vortices
across gradient.   These reversals
can occur as a function of vortex density,  ac driving amplitide, and pinning strength.
In general the transverse ratchet effect has a magnitude that is
about one-half to one-third the size of the longitudinal  
ratchet effect observed for the same conformal pinning arrays. 
We also examine the case where a dc drive 
applied perpendicular to the asymmetry of the substrate
produces what is known as a geometric or drift ratchet, where
a net flux of vortices perpendicular to the dc drive occurs.
The maximum efficiency of the drift ratchet 
is about 2.5 times larger than that of the 
corresponding ac driven transverse ratchet. 
The drift ratchet is also a realization of a noise correlation ratchet, 
and we find that it exhibits reversals in the drift
direction as a function of field, drive amplitude, and pinning strength.
Our results should be general to a wide class of systems
of interacting particles
undergoing dynamically generated fluctuations
when driven over a a conformal array, 
including colloidal particles in optical trap arrays.             

\section{Acknowledgments}
This work was carried out under the auspices of the 
NNSA of the 
U.S. DoE
at 
LANL
under Contract No.
DE-AC52-06NA25396.

\end{document}